\documentclass[structabstract]{aa}  
\usepackage[varg]{txfonts}
\usepackage{graphicx}
\usepackage{natbib}
\usepackage{amssymb}

\usepackage{etex}
\reserveinserts{18}
\usepackage{morefloats}

\begin{document}
\title{3D pyCloudy modelling of bipolar planetary nebulae: \\
evidence for fast fading of the lobes\thanks{Based on observations
  collected at the European Organisation for Astronomical Research in
  the Southern Hemisphere, Chile (proposal 075.D-0104) and HST
  (program 9356)} }

   \subtitle{}

   \author{K. Gesicki\inst{1} \and A. A. Zijlstra\inst{2}
           \and C. Morisset\inst{3} 
	  }

   \institute{Centre for Astronomy, 
              Faculty of Physics, Astronomy and Informatics,
              Nicolaus Copernicus University,
              Grudziadzka 5, 
	      PL-87-100 Torun, 
	      Poland, \email{kmgesicki@umk.pl}
              \and
              Jodrell Bank Centre for Astrophysics,
              School of Physics \&\ Astronomy,
	      University of Manchester,
              Oxford Road,
              Manchester M13\ 9PL, UK
               \and
              Instituto de Astronomía, Universidad Nacional Autónoma de México,
              Apdo. Postal 70264, Méx. D. F., 04510 México, Mexico
             }

\titlerunning{3D modelling of bipolar planetary nebulae: fast fading of the lobes}
\authorrunning{K.Gesicki et al.}


  \abstract
     {}
      {The origin and evolution of the shapes of bipolar planetary
        nebulae is poorly understood.  We postulate that their history can be traced
        through their internal velocity fields in a procedure
      similar to the one well established for spherical objects. 
      Such an analysis requires 3D photo-ionization and kinematical modelling which is
        computationally very time consuming.  We apply an axially
        symmetric pseudo-3D photoionization model, pyCloudy, to derive
        the structures of 6 bipolar
        nebulae and 2 suggested post-bipolars in a quest to 
      constrain the bipolar planetary nebulae evolution.  }
      {HST images and VLT/UVES spectroscopy are used
      for the modelling. The targets are located
      in the direction of the Galactic bulge. 
      A 3D model structure is used as input to the
      photoionization code, so as to fit the HST images. Line profiles of
      different ions constrain the velocity field. The model and associated
      velocity fields allow us to derive masses, velocities, and ages.  }
{The 3D models find much lower ionized masses than required in 1D models:
  ionized masses are reduced by factors of 2--7. 
  The selected bi-lobed planetary nebulae show a narrow range of ages: 
  the averaged radii and velocities result in values between 1300 and 2000\,yr.
  The lobes are fitted well with velocities linearly increasing with radius.
  These Hubble-type flows have been found before,
  and suggest that the lobes form at a defined point in time. 
  The lobes appear to be slightly younger than the main (host)
  nebulae, by $\sim 500$\,yr, they seem to form at an early phase of PN
  evolution, and fade after 1--2\,kyr. We find that 30--35\%\ of bulge PNe
  pass through a bipolar phase.  }
     {}

   \maketitle
%

\section{Introduction}

\begin{figure*}
\centering
 \includegraphics[height=4.2cm]{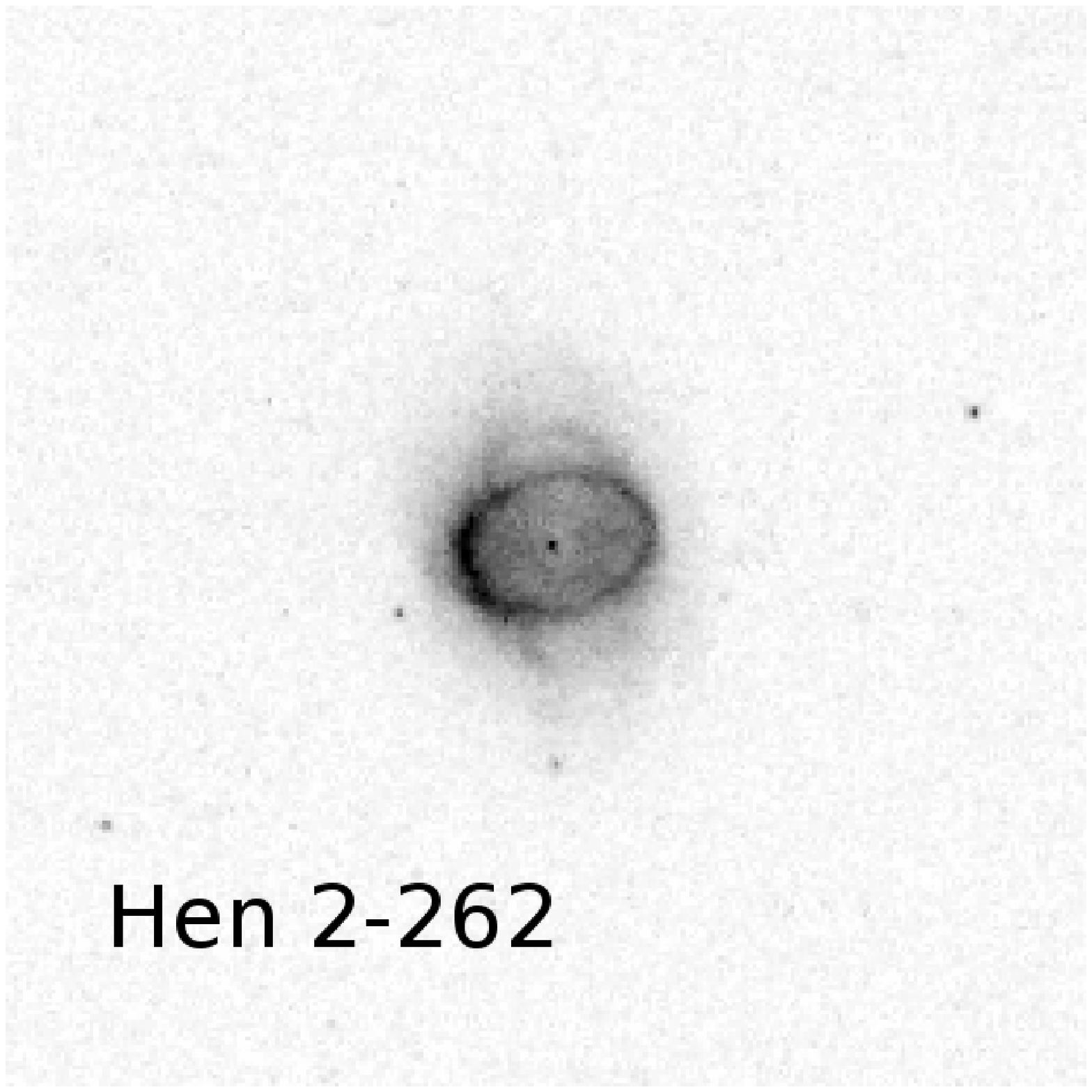}
 \includegraphics[height=4.2cm]{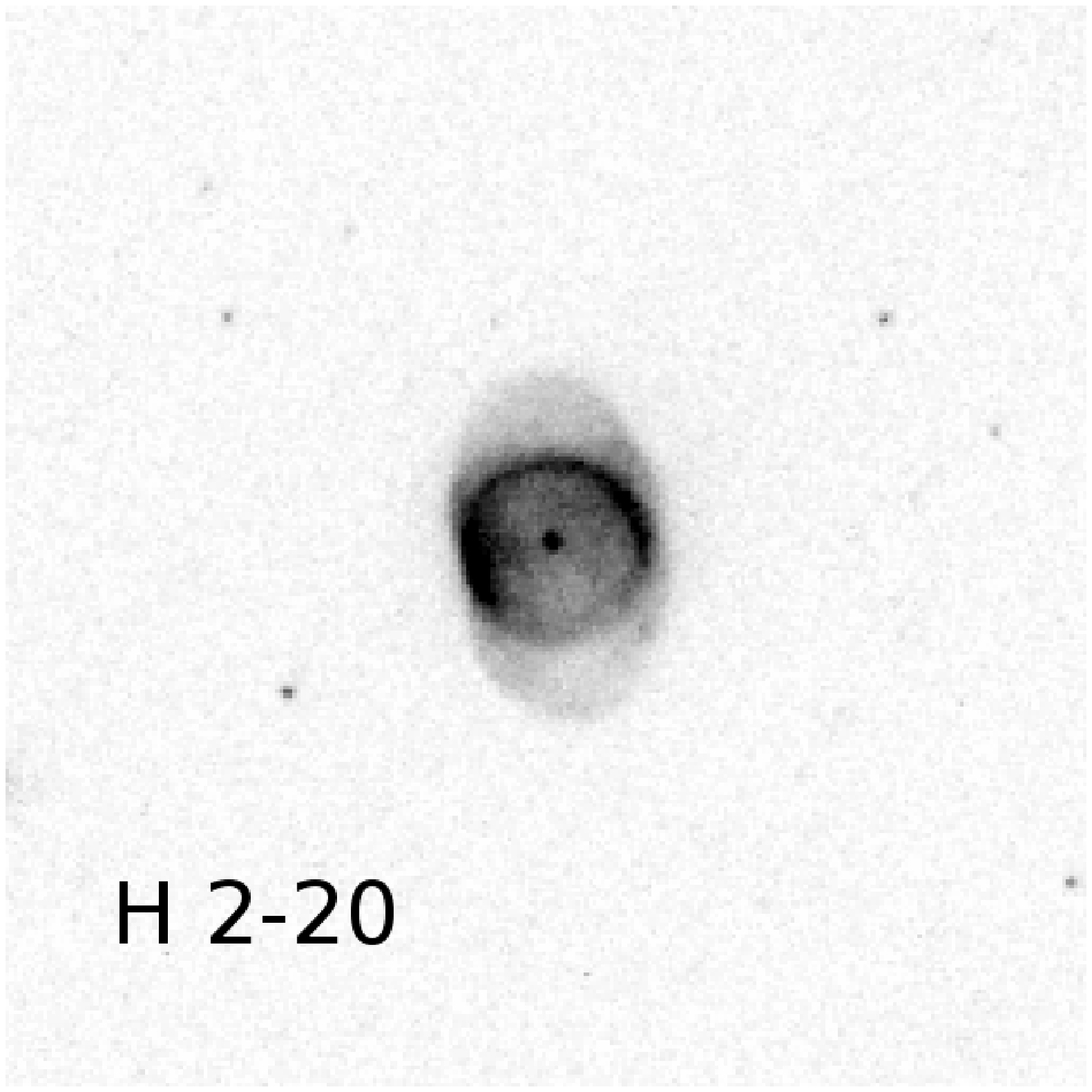} 
 \includegraphics[height=4.2cm]{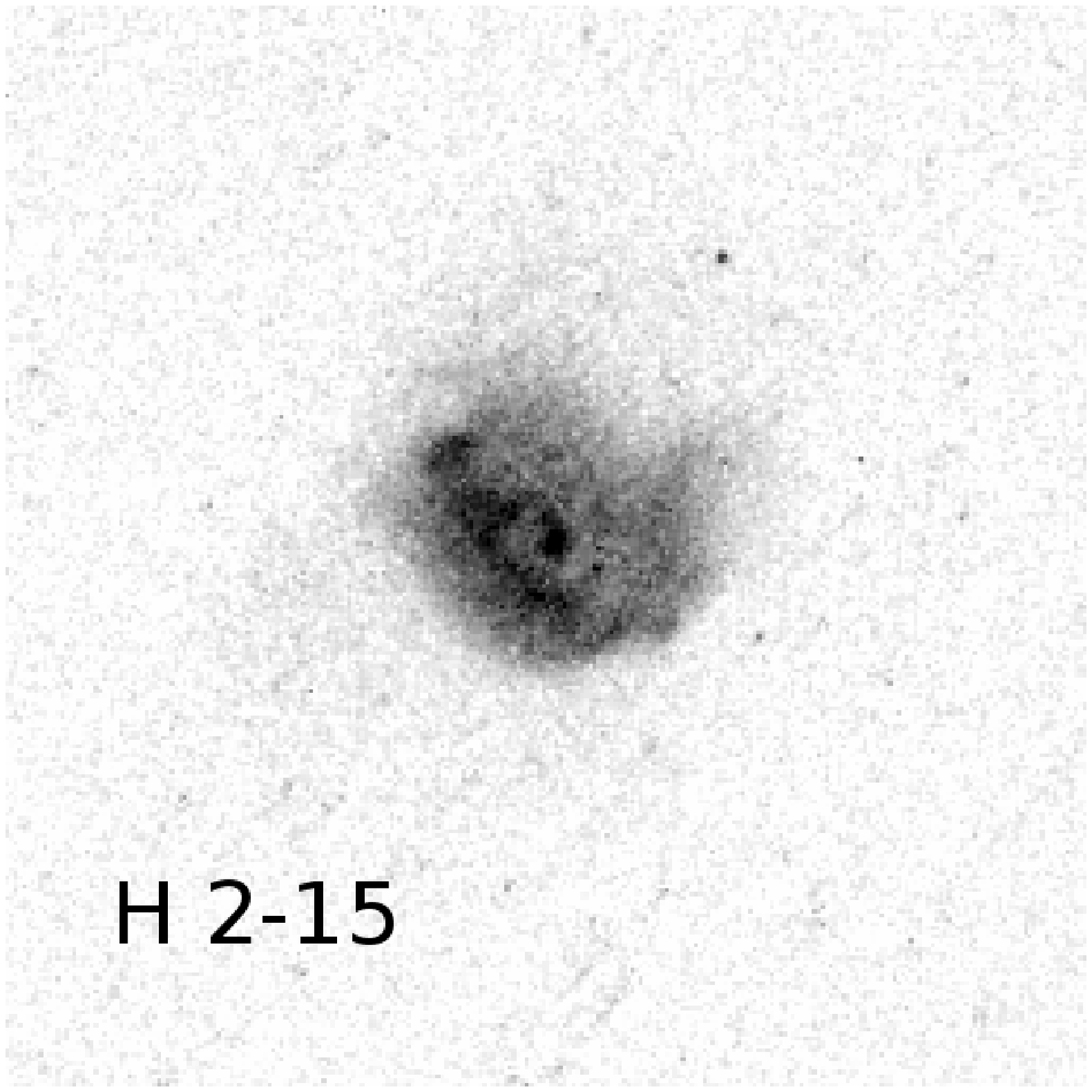} 
 \includegraphics[height=4.2cm]{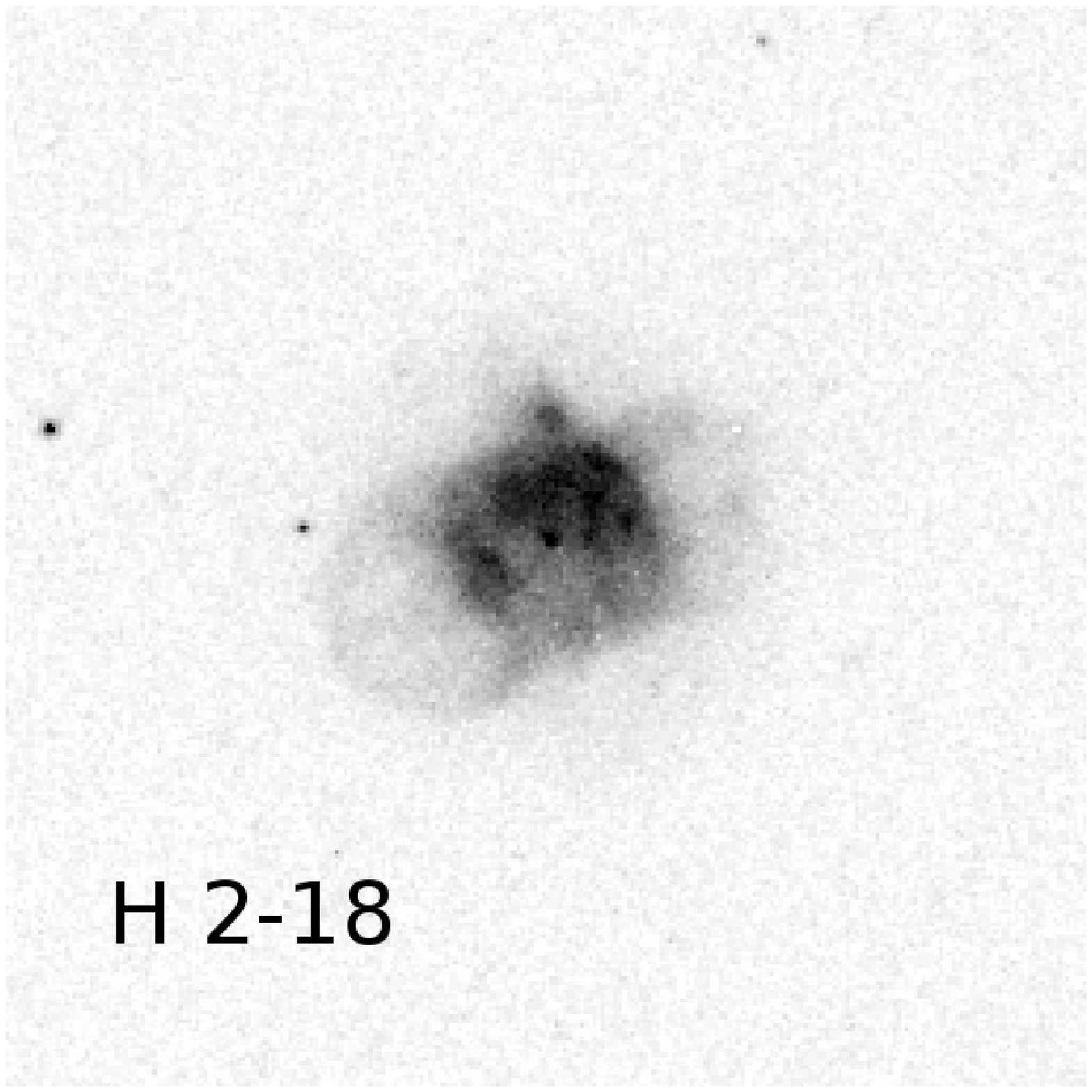}
 \includegraphics[height=4.2cm]{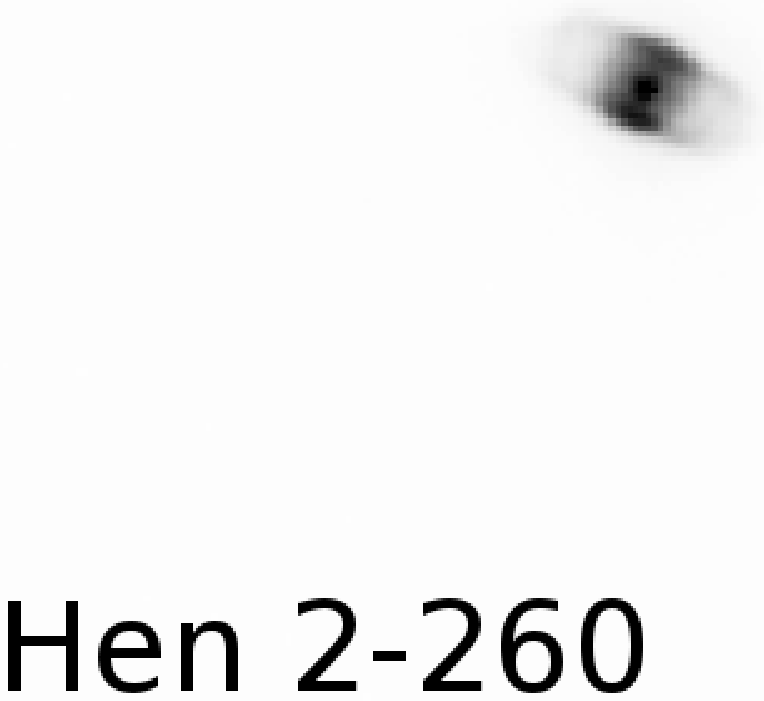}
 \includegraphics[height=4.2cm]{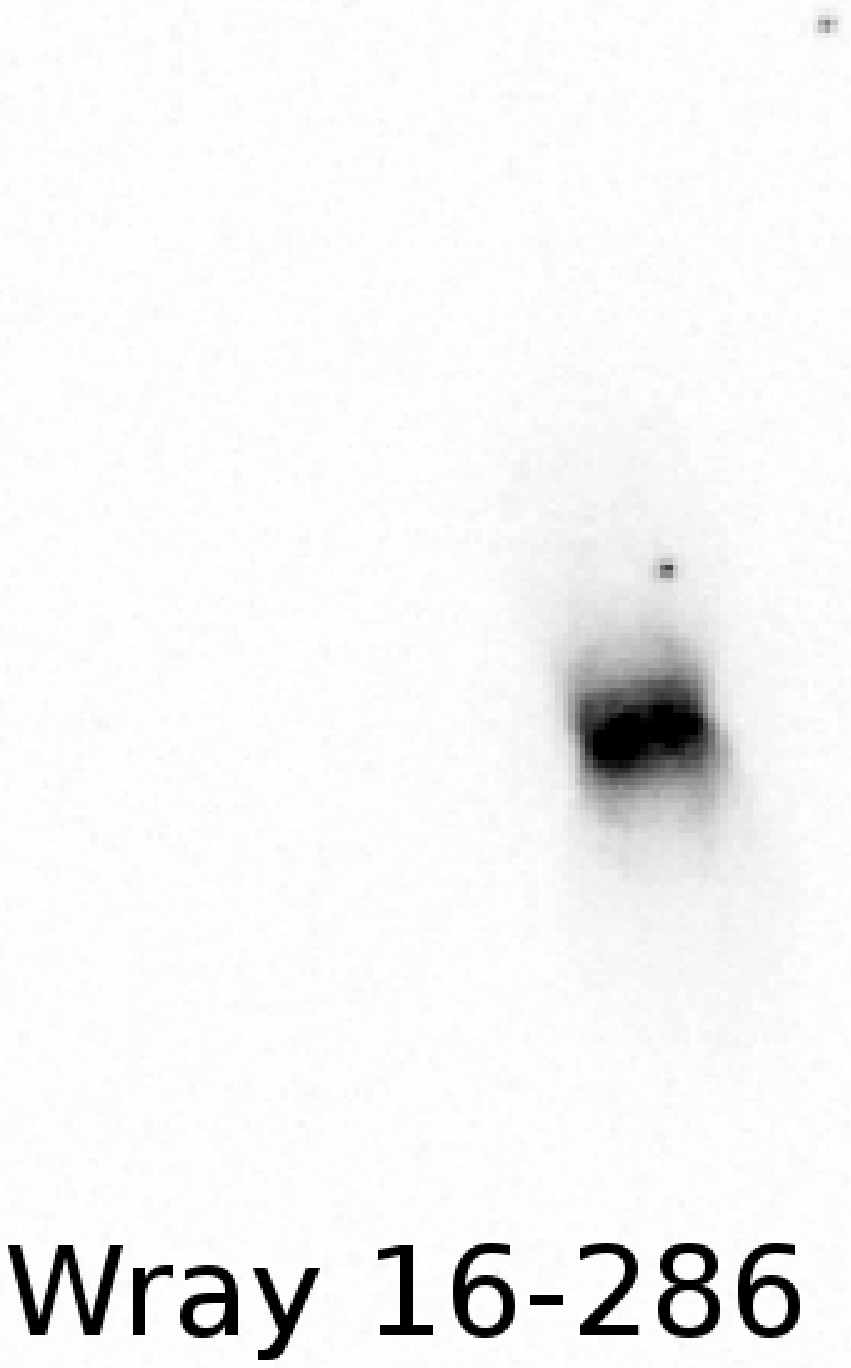} 
 \includegraphics[height=4.2cm]{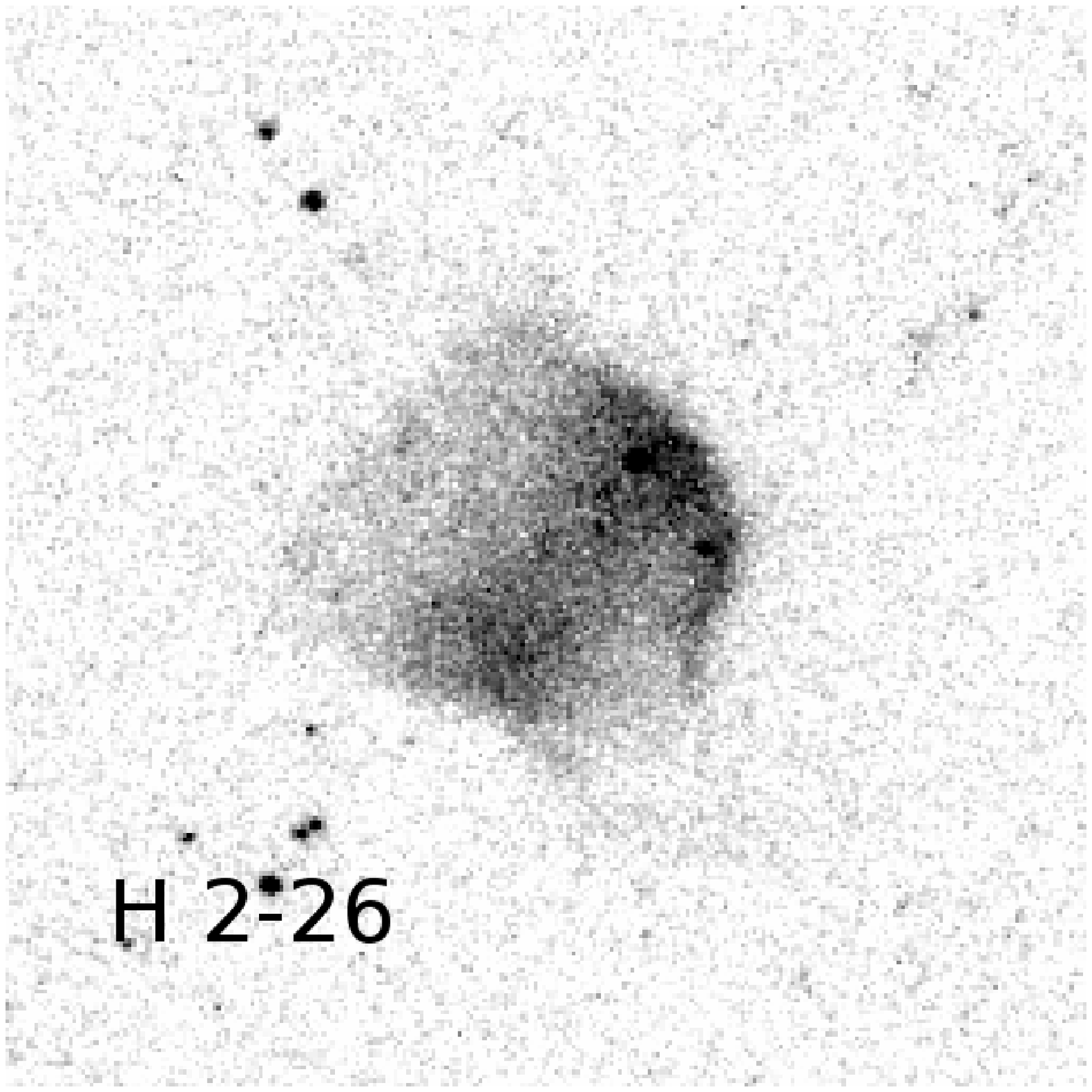}
 \includegraphics[height=4.2cm]{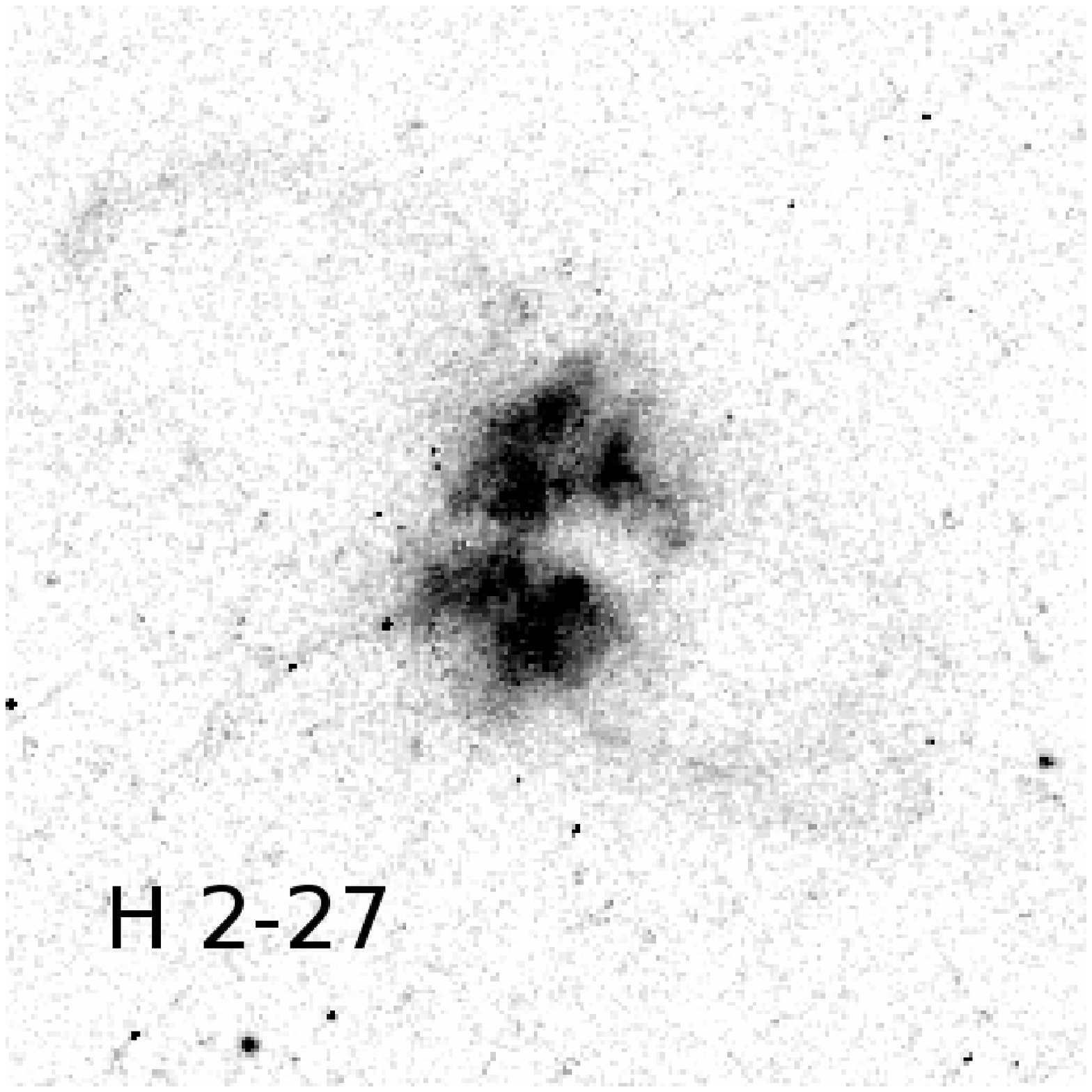}
 \caption{H$\alpha$ HST images presented to the same angular
   scale. Each panel is 12\,arcsec on the side. The intensity scale is
   linear. The names are indicated in each panel.  The first three
   columns show PNe in the bipolar phase while the rightmost column
   shows the proposed post-bipolar phase, discussed in the text. Six
   PNe are assumed to belong to the Galactic bulge while the smallest
   and largest, both shown in the third column, are assumed to be
   located behind and in front of the bulge, respectively.}
 \label{hst_ima}
\end{figure*}

A planetary nebula (PN) forms when a low- or intermediate-mass star at
the end of its life ejects much of its envelope into space.  The nebula
initially expands with the ejection velocity.  As the central star heats up,
it ionizes the nebula, and the resulting ionization front and the
over-pressure of the ionized region alter the velocity field and density
structure.  Interaction with the fast wind from the hot central star causes
further changes.  The velocity field therefore contains information both on
the origin and the evolution of PNe.

For spherically symmetric nebula, these interactions have been modelled by
the Potsdam group, e.g.  \citet{jaco2013, scho2014} and references therein. 
These articles discuss a relation between the true nebular expansion and
differently estimated expansion velocities.  The different kinematical ages
are compared with the true post-AGB age.  \citet{scho2014} and
\citet{gesi2014} obtained quite independently the same result that for more
or less regular PNe, the true post-AGB age and kinematical age obtained from
the mass-averaged expansion velocity are in agreement to a correction factor
of 1.4.  The `true' post-AGB age, in both papers, is the elapsed time since
the star left the AGB, as obtained from the evolutionary computations of
\citet{bloc1995}.

Observationally, the velocity fields can be derived from the line profiles
of different ions, where each ion traces a different layer due to the
ionization stratification.  Kinematical reconstruction of line profiles is
done by fitting a photoionization model to the observed nebular parameters
and then finding the velocity field which provides the best agreement
between the calculated and the observed emission line profiles.  The
suitably averaged velocity and the outer radius result in kinematical age. 
Such methods have been used successfully for spherical models of PNe in
\citet{gesi2006}, using 1-dimensional photo-ionization models.  Using HST
imaging combined with VLT spectroscopy, \citet{gesi2014} studied the
velocity fields of a sample of 36 compact PNe in the Galactic bulge
direction.  This also used the 1-dimensional Torun photo-ionization codes,
and thus implicitly assumed spherical symmetry.

However, most planetary nebulae deviate from spherical symmetry.  

The evolution of bipolar PNe is quite successfully described in terms
of the so-called ``generalized interacting stellar winds'' model.
Already \citet{bali1987} schematically outlined the proposed
morphological sequences of such evolution. Recently \citet{huar2012},
using extensive hydrodynamical simulations, studied the morphological
evolution of bipolar nebulae. More complex interactions, such as the
wind reflected of a warped disk, allows one to generate more complex
nebular shapes \citep{rijk2005}. However full 3D radiation-hydrodynamic
calculations are very time-consuming, and completed evolutionary
sequences, like those of Potsdam group for 1D, are not available yet.

In this paper for the first time we will study the
multi-directional structure and velocity field, for bipolar PNe.  We
will reconstruct the images and the emission line profiles following
a procedure similar to the one elaborated for spherical objects
which is suitable for modelling of a larger sample.  Having derived
H$\beta$-weighted averaged velocities we will estimate the
kinematical ages.

For the kinematical reconstruction we
will use a concept of pseudo-3D models described in \citet{moris2005}. 
The basic idea is to compute a grid of 1D photoionization models 
with each model corresponding to the given direction ($\theta$, $\phi$) 
pointing away from the centre of the nebula. A 3D cube of coordinates is build, 
on which interpolation of any physical parameter is made based on the 1D runs: 
for each cell of coordinates ($r$, $\theta$, $\phi$), the program looks for 
the closest three 1D-models that enclose $\theta$ and $\phi$, and interpolates 
on the radii to find the value corresponding to the given cell. 
In the case of an axially symetric nebula, there is no need for interpolation 
on the longitudinal angle $\phi$, only the latitudinal angle $\theta$ is used. 
These models are as accurate as full 3D as long as the non-radial 
radiation does not dominate the ionization (this would happen in shadowed 
regions ionized by radiation from the surrounding material or in a nebula 
ionized by multiple sources, which are not the cases here). 

After a short presentation of the observational material we will describe
the applied \emph{pyCloudy} python library and the derived with it pseudo-3D models. 
For each of the selected objects the best fitted model
will be presented in separate figures and subsections.  Next we will search
for correlations between the new parameters and discuss them.  Eventually we
propose an evolutionary scenario in which the lobes evolve surprisingly fast
-- the obtained timescales for the bipolar phase are an order of magnitude
smaller than the often assumed visibility times \citep{jaco2013}.

\section{Observations and analysis}
\label{obse_anal}

\subsection{The selected images and spectra}

The observational data are presented in \citet{gesi2014} and described
there in more detail. The images of compact bulge PNe were obtained
with HST/WFPC2, in H$\alpha$ and [\ion{O}{iii}]\,5007\,\AA. For each object a
series of 600-second VLT/UVES echelle spectra were obtained, with the
slit centred on the nebula and aligned in most cases with the minor
axis.

The sample of 36 bulge PNe was randomly selected from compact PNe listed in
the SE\,CGPN \citep{acker1992}, with sizes of 5 arcsec or less, and is not
biased towards any morphology nor brightness.  Among these PNe there are
some 20 objects composed of a central structure (ellipsoid or disk) plus a
clear extension.  Those extensions can be of the shape of ``classic''
bipolar lobes (6 PNe), multipolar (quadrupolar or more) lobes (10 PNe),
curved wing-like open structures (5 PNe).  Some further objects are
ellipsoidal with or without inner structures or completely irregular.  (A
detailed morphological classification can be found in \citet{sahai2011}.)

From this observed sample, for the 3D analysis we selected objects which
appear similar and exhibit both axial-symmetry and point-symmetry.  These
PNe show both a central structure (which can be interpreted as an equatorial
torus) and a pair of rather wide, extended and closed ``classic'' lobes. 
Both lobes are the same and oriented perpendicular to the equatorial plane. 
This criterion is fulfilled by six objects only: Hen\,2-262, H\,2-20,
H\,2-18, Hen\,2-260, Wray\,16-286, and H\,2-27.  It is not unlikely that
some other objects in the complete sample should fall into this group. 
Round PNe could be bipolars seen along the symmetry axis.  However for
clarity of this analysis we did not consider these doubtful cases.  We added
two more objects to the present work: H\,2-15 and H\,2-26.  Because of
their morphology and their old age assigned by \citet{gesi2014} we suggest
that both are in a post-bipolar phase, and thus provide a useful comparison
sample discussed later in Sect.\,\ref{postbi1} and Sect.\,\ref{postbi}.

Fig.\,\ref{hst_ima} presents the H$\alpha$ HST images. The intensity is
represented as a linear grey scale. The images reveal some details of
the bipolar lobes; the lobes are much weaker in intensity than the
torus. The angular size of each square is the same (12\,arcsec on the
side) to visualize the differences in angular size between the PNe. 
The two proposed post-bipolar PNe are shown in the Fig.\,\ref{hst_ima} in
the right-most column.

The selected emission lines are shown in the right columns of
Figs.\,\ref{mode0012}--\ref{mode3561} together with the fitted models.
The spatial resolution of the UVES data is much lower than that of
HST:  the slit was 0.5\,arcsec wide and the seeing was about
1\,arcsec (typically between 0.6 and 1.4 arcsec, for H\,2-27 even
above 2.5 arcsec).  For the analysis of the line profiles, the
central arcsecond was extracted from the observed slit
spectra. Thus, the analysed line profiles are those observed near
the centre of each nebula. This suffices to derive the velocity of
the inner regions, but the far-away tips of the lobes lay beyond the
long-slit so that the extremum velocity is not as well constrained
by the observations.

\subsection{pyCloudy and 3D photoionization modelling}

The \emph{pyCloudy} code \citep{moris2013} used for the modelling is
available on a web
page\footnote{https://sites.google.com/site/pycloudy/} together with
documentation and instructions.  It consists of a photo-ionization code
(for which \emph{Cloudy}\footnote{http://www.nublado.org/}
\citep{Ferland2013} is used) and a python library that contains
routines to analyse the models. Included are routines to visualize the
models, to compute the (differently defined) averaged expansion
velocities, to integrate the masses, fluxes etc.

As already described this method of solving for the structure is
pseudo-3D rather than full 3D. The python codes in their current
version limit the modelling to axial- and point-symmetric objects. 
The photo-ionization models are themselves 1D, and therefore do not
consider correctly the effect of the diffuse radiation field. The
interpolation introduces further uncertainties. It is however
computationally fast. Full models, e.g. using Monte Carlo methods 
(see for example \citet{dane2013}), are
computationally very expensive. In the current work we will explore
whether the data can be fitted with the pseudo-3D models. Full 3D
models are left for later.

We will use a blackbody spectrum for the central star, as was done in
previous 1D models. These models tend to return a lower stellar
temperature than using model atmospheres, because they lack the Lyman
jump. We ran several tests comparing blackbody SED with the
stellar atmospheres SED available at Cloudy web page, in particular
with the PN central star models of \citet{rauch2003} and found that
this had no significant effect on the velocity fields and ages. The
adopted blackbody temperatures $T_\mathrm{bb}$ are listed in
Table\,\ref{param_photoio}.

\subsection{Density structure}

The photoionization part of pyCloudy computes a sequence of
one-dimensional regular Cloudy models aimed to represent the
structure of a single nebula diagnosed at many different radial
directions. Each 1D model actually represents a spherically
symmetric PN and consists of an empty inner region extending from
the centre up to the specified inner radius, then follows the main
nebular body at a constant density, and terminates at the specified
outer radius or at the ionization boundary, whichever comes
first. Produced are 1D grids of line emissivities, electron
temperatures, etc.

Each computed 1D model corresponds to a selected radial
direction, inclined to the equatorial plane at a specified
(latitudinal) angle with varying inner and outer radii.  In our
models, the first is at $0\degr$ (equator) and the last at $90\degr$
(symmetry axis). The angles in between are selected based on the
structure of each nebula, and are different for each model. The
spacings do not need to be uniform in angle. We used 8 separate
angles, based in part on the required computational time. A test run
with a set of 16 angles showed little improvement, and for our
structures sets of 8 angles appear to be sufficient.

pyCloudy re-grids the obtained 1D distributions into a XZ
cross-section of a 3D cube.  Because of the introduced symmetry
centre and symmetry axis only one quadrant of the plane needs to be
filled.  The space between the 1D lines is filled in by
interpolation while the remaining quadrants of the cross-secting
plane are filled in by symmetry conditions.  These planes are shown
for each analysed object in Figs.\,\ref{mode0012}--\ref{mode3561} in
the upper-left panels.  The symmetry axis is placed horizontally.
The gas density is shown in grey shades in linear scale but plotted
only for the ionized part of the nebula.  In one quadrant the
(magenta) line segments indicate the eight directions along which
the 1D Cloudy models were computed. The outer end of each segment
corresponds to the assumed end of gas distribution i.e. the matter
(density) boundary.  Wherever the shown density is not reaching the
line-segment's end this means that in this direction the actual
nebular edge is defined by the ionization front. 

The full cube is filled by pyCloudy through rotation around the
symmetry axis.  To fill the cross-secting plane and the full cube
the standard linear interpolation of python was applied.  Routines
are available to compute the projection of the full cube onto the
plane of the sky, defined by the inclination angle between the line of
sight and the symmetry axis. In this way model images can be obtained
for any of the calculated emission lines. For H$\alpha$ they are shown
in Figs.\,\ref{mode0012}--\ref{mode3561} in the centre-left panels.
Very different nebular structures can be obtained through varying the
shapes of the inner voids and outer lobes.

The computed images are compared visually with the HST images
(shown in Figs.\,\ref{mode0012}--\ref{mode3561} in the bottom-left
panels) and then by trial and error the inner and outer radii, the
inclination of the symmetry axis, the densities and latitudinal
angles of the individual sectors are optimized.  Their
numerical values are given in Table\,\ref{param_geom}.

\subsection{Velocity field}
\label{veloc}

A velocity field is defined where all velocities are directed
radially with respect to the central star, and the velocity only
depends on the radial distance from the centre. Note that this is
simpler than the density structure: the material has different
spatial distribution at different latitudinal angles while the
velocity distribution is the same for any inclination. More complex
velocity fields (depending e.g. on the angular distance from the
symmetry axis) are possible but go beyond the scope of this paper,
or the observational constraints.

A constant expansion velocity is inconsistent with the different
widths of lines of different ions, which indicate higher velocities in
the outer (low ionization) regions.  To limit the number of the model
free parameters, the velocity was restricted to be linearly increasing
with radial distance, starting from zero at the central star and
reaching a maximum at the ends of lobes. This  Hubble-type
  dependency (alternatively often called homologous) improved the
model. Trial computations showed a need to further modify this simple
scheme.  We introduced an inner region with a constant velocity.
Thus, the adopted velocity fields are composed of an inner plateau
(its velocity and radial extent was adopted by trial and error, guided
by the [\ion{O}{iii}]\,5007\,\AA\ line profile), connected to the
linearly increasing part further out (guided by the
[\ion{N}{ii}]\,6583\,\AA\ line).  The spherical symmetry of the
velocity field combined with axial and point symmetry of the density
distribution provides a slow expansion for the dense regions close to
the star (dominated by the equatorial torus at low inclination angles)
and fast expansion for the far-away extended and thin lobes at high
inclination angles.

We stopped at this level of complication. Such a situation appears
reasonable for 3D bipolar models and corresponds quite well to many of
the 1D models (composed of a dense inner shell and an extended outer
halo e.g. \citet{gesi2014}). Such velocity is in agreement with ``a
typical middle-aged and fully ionized nebular structure'' obtained
with the 1D hydrodynamical modelling of \citet{scho2014}.  Even better
agreement is with the 2.5D hydrodynamical calculations of
\citet{huar2012} where some of the models show a Hubble-type expansion
combined with closed bipolar lobes.

Our velocity field is also similar to those found observationally
for bi-lobed nebulae.  For the well-resolved nebula NGC\,2346 at a
distance of 700\,pc, \citet{aria2001} derived a best model with
``outflow velocity directed radially and proportional to the radial
distance''.  The NGC\,2346 expansion velocities of 16\,km\,s$^{-1}$
at the central ring and of about 60\,km\,s$^{-1}$ at the tips of the
lobes are comparable to many of our values.  For two other nebulae,
NGC\,6302 \citep{szys2011} and NGC\,7027 \citep{walsh1997} also
similar velocity fields were found although both objects have a
rather dramatic appearance. \citet{alco2007} analysed molecular
gas in the pre-planetary nebula M\,1-92 (Minkowski's footprint) and
found a Hubble-like velocity field not only in bipolar flow but also
in the equatorial plane. Note that this situation is neither new
nor unique and is typical to our whole sample.

Line profiles can be calculated by integrating the particular line of sight
through the cube, at any position within the projected image. Summing
these over a defined area simulates the line profile obtained through a
slit. The spectrograph slit is assumed in the shape of a rectangle
around 1.5\,arcsec on the side. This includes the effect of seeing.
Positioned centrally on the nebular image, this produces symmetric line
profiles. However when the rectangle is shifted somewhat off-centre,
calculations show an asymmetric profile. This effect is well known, see
e.g. \citet{mont2000}. The asymmetry obtained in this way fits quite
well to the observed spectra and this was done for some of our objects.
In Figs.\,\ref{mode0012}--\ref{mode3561} in the middle-left panels 
we indicate with the (red) rectangle the applied size of the slit
superposed on the projected nebula image. The right panels show the computed
line profiles (dashed red lines) obtained with this setup.

However, there is an ambiguity: we cannot tell whether the observed
asymmetry comes from a slight offset of the spectrograph slit (which
for extended weak objects is very likely) or from real irregularity
within the nebulae themselves (which the images indicate is also
present in some cases).

\subsection{Inclination angle}

The derivation of geometrical dimensions and of the expansion
velocities for bipolar PNe is much more complicated than for
spherical ones. The angle of inclination between the polar axis and
the plane of the sky affects several parameters and requires 
special attention.

\begin{figure}
\centering
\includegraphics[width=6cm]{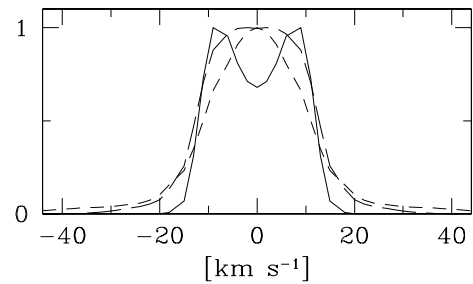}
\caption{Emission line profiles for different inclination angle. 
Plotted is the [\ion{N}{ii}]\,6583\,\AA\ line computed for the model of
the spatially unresolved nebula Hen\,2-260. The solid line represents
the inclination angle of $0\degr$ (symmetry axis within the plane of the sky),
long-dash line -- inclination of  $30\degr$, short-dash -- of $60\degr$. }
 \label{a_rot}
\end{figure}

The dominant part of the line emission originates in the dense nebular
torus. If the symmetry axis lies in the plane of the sky then most of
nebular material moves towards to and away from the observer and the
resulting emission profile is clearly split, with a dip at the
systemic velocity. At the other extreme, when the torus lies in the
plane of the sky the bulk expansion proceeds perpendicularly to the
line of sight, the line narrows and the splitting disappears. The
observed splitting of the brightest part of the line profiles can be
used to derive the best-fitting inclination of the symmetry axis of
the model as illustrated in Fig.\,\ref{a_rot}. This figure was
plotted for [\ion{N}{ii}]\,6583\,\AA\ line of Hen\,2-260 for
inclination angles of $0\degr$, $30\degr$ and $60\degr$. This plot
can be compared with Fig.\,\ref{mode0082} bottom-right panel where
an inclination of $50\degr$ was adopted. The nebula is spatially
unresolved by the spectrograph therefore the behaviour shown in
Fig.\,\ref{a_rot} does not depend on the selection resulting from an
applied small slit area. This kind of analysis is possible thanks to
the high spectral resolution (about 5\,km\,s$^{-1}$) and this
example shows that the accuracy of the inclination of such a
procedure can be roughly estimated as $\pm 10\degr$.

However the splitting depends also on the assumed velocity
field. Further, the derived inclination angle determines the true
extent of the model which is used in the velocity field.

Another effect can be deduced from Fig.\,\ref{a_rot}: with
increased inclination angle the low-intensity and high-velocity line
wings become apparent.  This follows from the fact that for the
highly inclined object the extended fast lobes expand in the
direction closer to the line-of-sight and the emission becomes
Doppler-shifted.  Because of this effect the lobes are partially
available for the kinematical reconstruction even despite the
limited spectrograph slit area. For the angles of about
$40\degr-50\degr$ quite a significant part of the velocity field can
be probed. However for smaller inclination angles and larger nebulae
the lobe's velocity has to be extrapolated with the assumed
homologous expansion law derived based on the central regions.
Certainly `integral field unit' spectroscopy would provide data much
better suited for this kind of analysis; nevertheless even this would
not help to measure the gas motions in the plane of the sky. 

All these factors introduce a type of degeneracy. The degeneracy is
reduced by the restriction to linearly increasing velocity fields and
ellipsoidal shapes. Careful examination of details of the images and
spectra assures us that the model is a good proxy for the real nebulae
but each solution may not be unique.

\section{pyCloudy models for individual objects}
\label{phi_mod}

\begin{figure}
\resizebox{\hsize}{!}{
 \includegraphics[height=7cm]{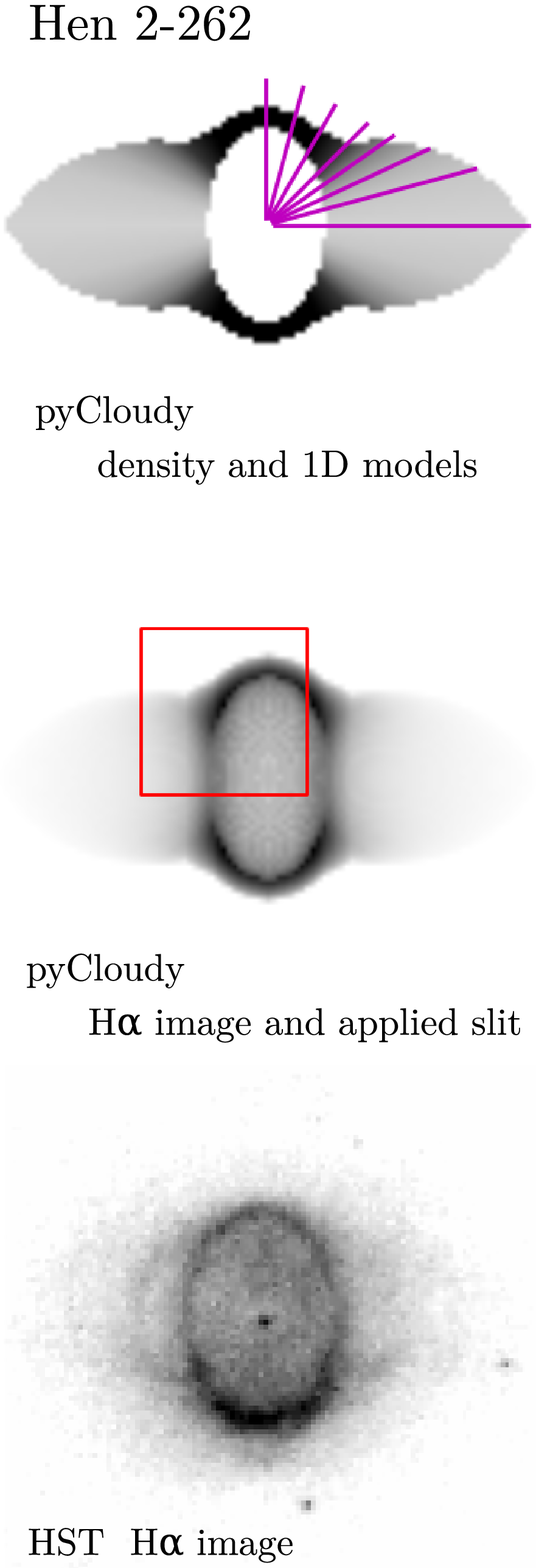}
 \includegraphics[height=7cm]{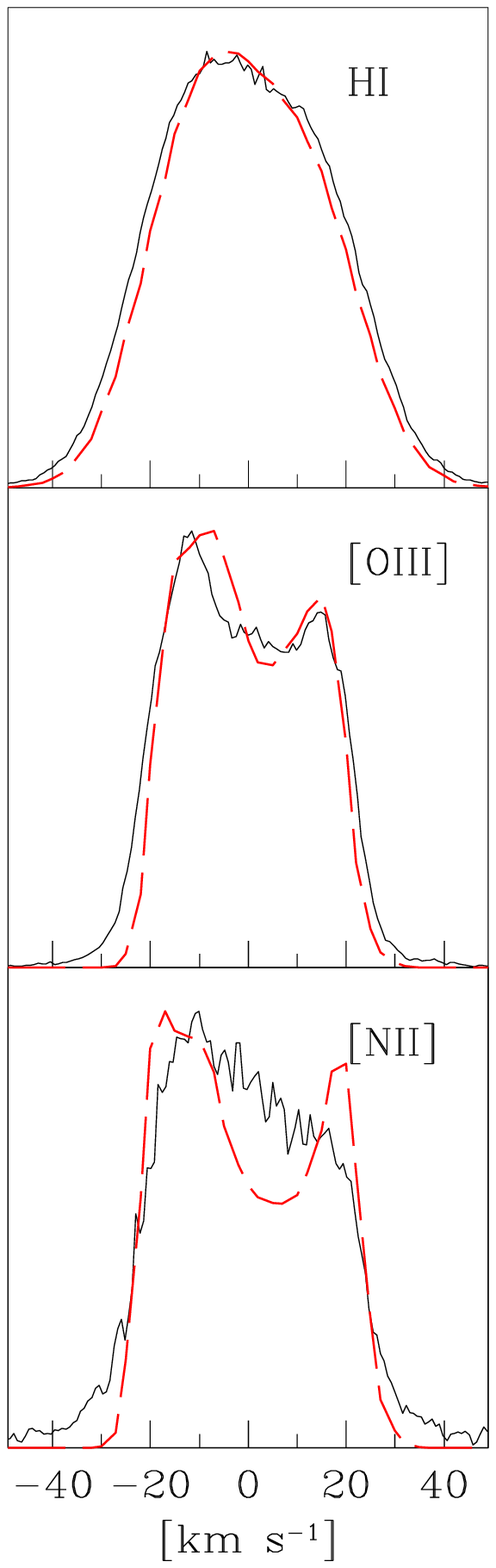} }
\caption{Hen\,2-262.  The top left panel shows the assumed
    density structure for the ionized nebula only, drawn at the
    cross-section plane taken along the positioned horizontally
    symmetry axis.  The (magenta) lines indicate the directions along
    which the 1D constant density Cloudy models were calculated. The
    end of each line segment corresponds to the assumed matter
    (density) boundary. The middle-left panel shows the H$\alpha$
  image (i.e. a projection onto a plane of the 3D structure) obtained
  for the same model but inclined to the sky plane at $20\degr$ to
  correspond approximately to the HST image, shown in the bottom left
  panel (also in Fig.\,\ref{hst_ima}).  The intensity grey scale is
  linear in all three images, the size is scaled to fill the width of
  the box.  The (red) rectangle in the middle panel indicates the
    area over which the line profiles are integrated. The nebular
  emission line profiles are shown on the right, the solid (black)
  lines show the observations while the dashed (red) lines show the
  computed model, for H$\alpha$, [\ion{O}{iii}] 5007\,\AA, and
  [\ion{N}{ii}] 6583\,\AA.  }
 \label{mode0012}
\end{figure}

\begin{figure}
\resizebox{\hsize}{!}{
 \includegraphics[height=7cm]{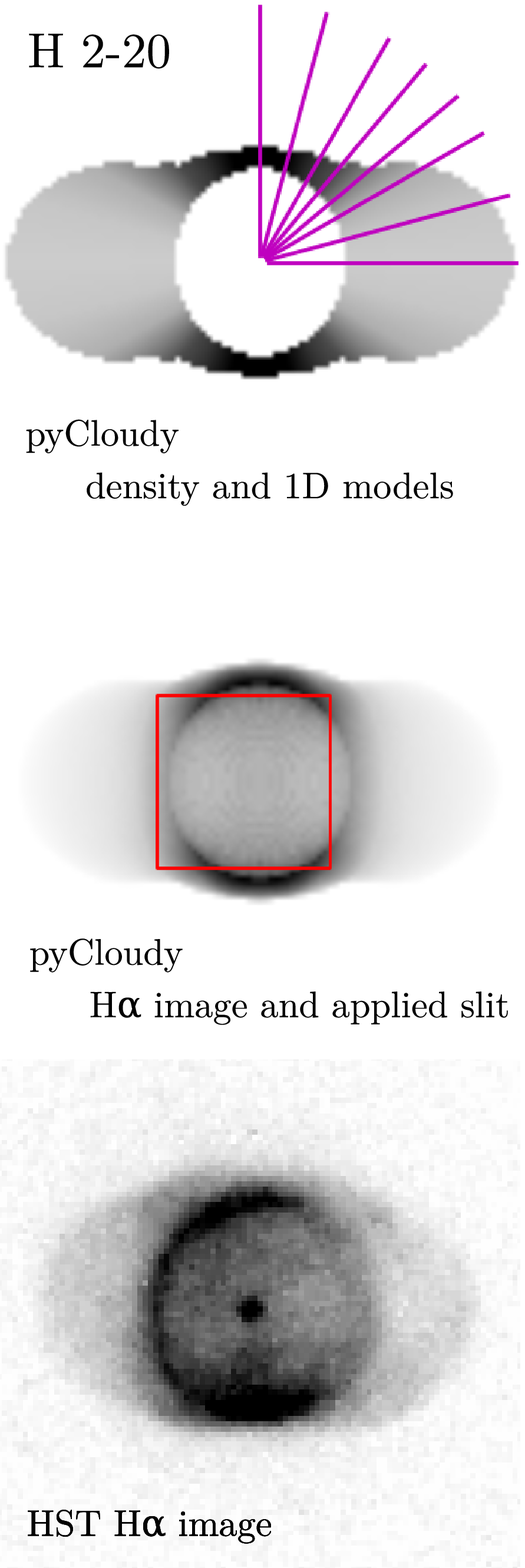}
 \includegraphics[height=7cm]{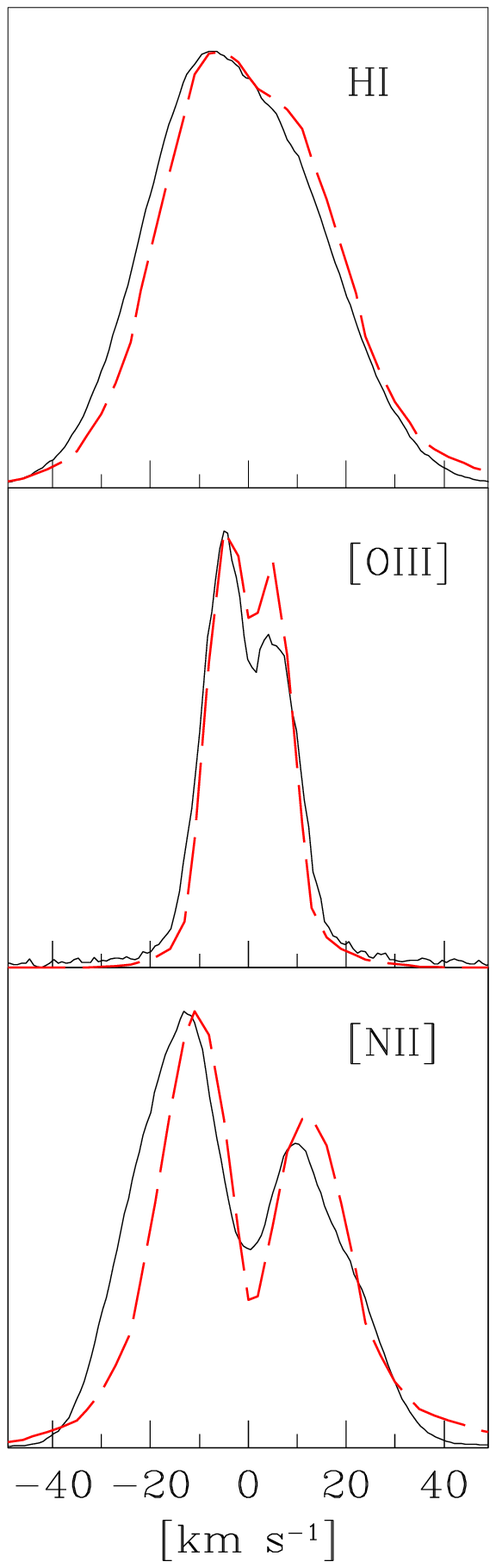} }
 \caption{H\,2-20. Data presented as in
   Fig.\,\ref{mode0012}. The axis inclination is $30\degr$.}
 \label{mode0028}
\end{figure}

\begin{figure}
\resizebox{\hsize}{!}{
 \includegraphics[height=7cm]{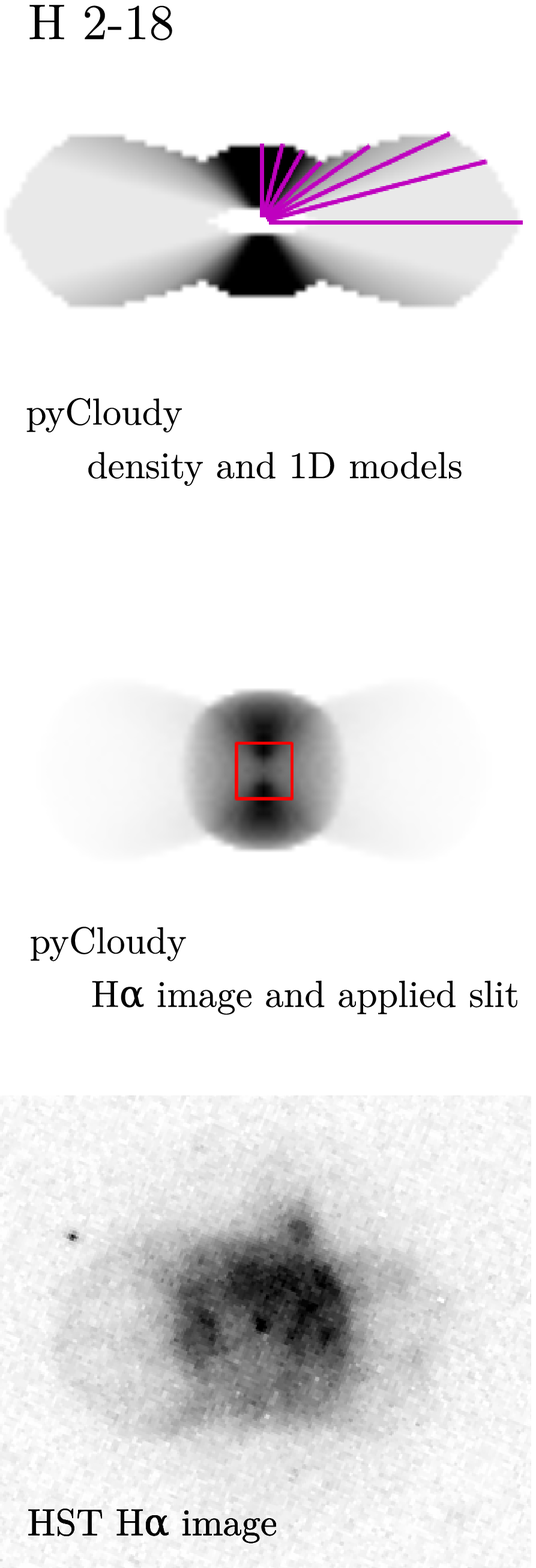}
 \includegraphics[height=7cm]{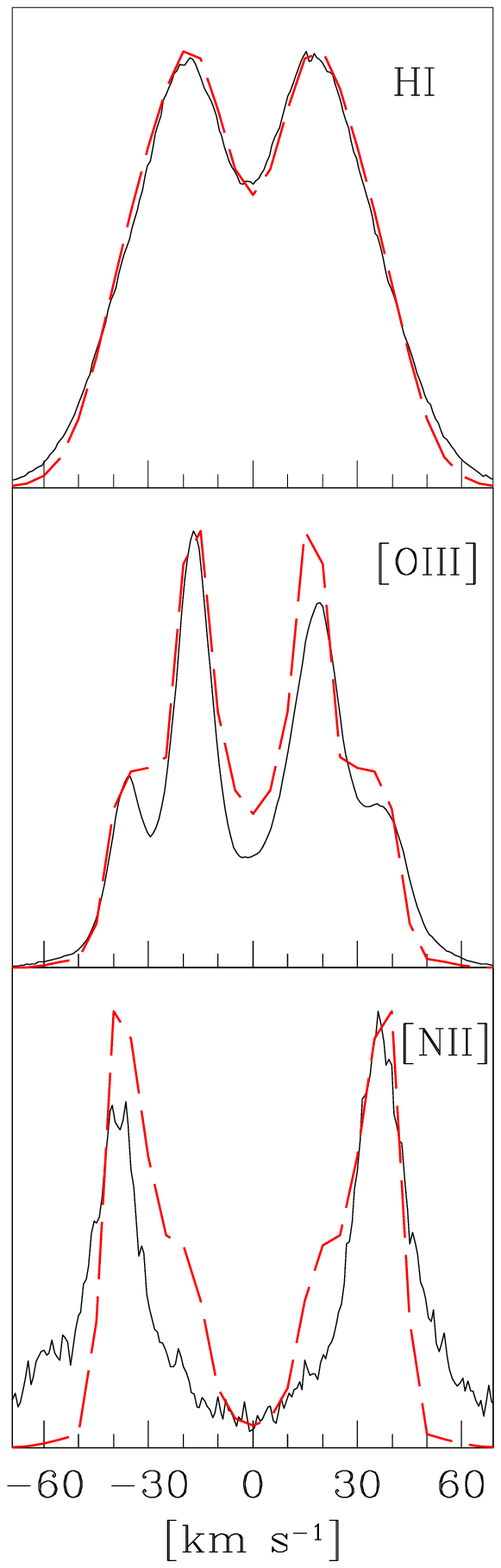} }
 \caption{H\,2-18.  Data presented as in
   Fig.\,\ref{mode0012}. The axis inclination is $35\degr$. }
 \label{mode0063}
\end{figure}

\begin{figure}
\resizebox{\hsize}{!}{
 \includegraphics[height=7cm]{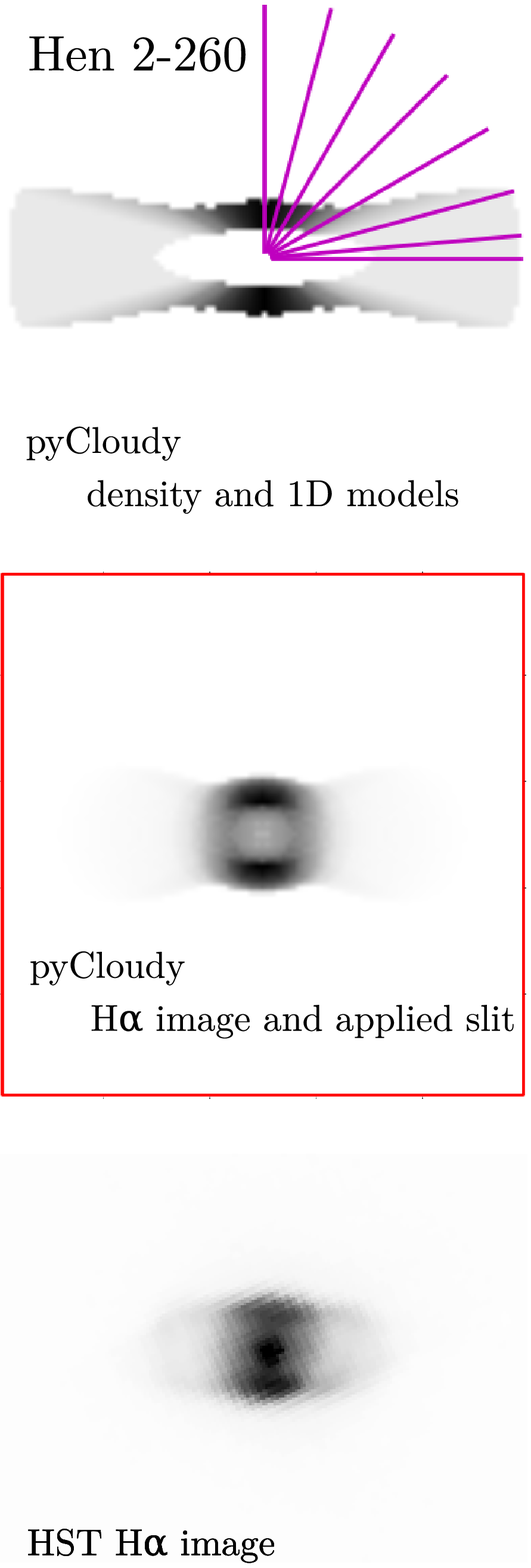}
 \includegraphics[height=7cm]{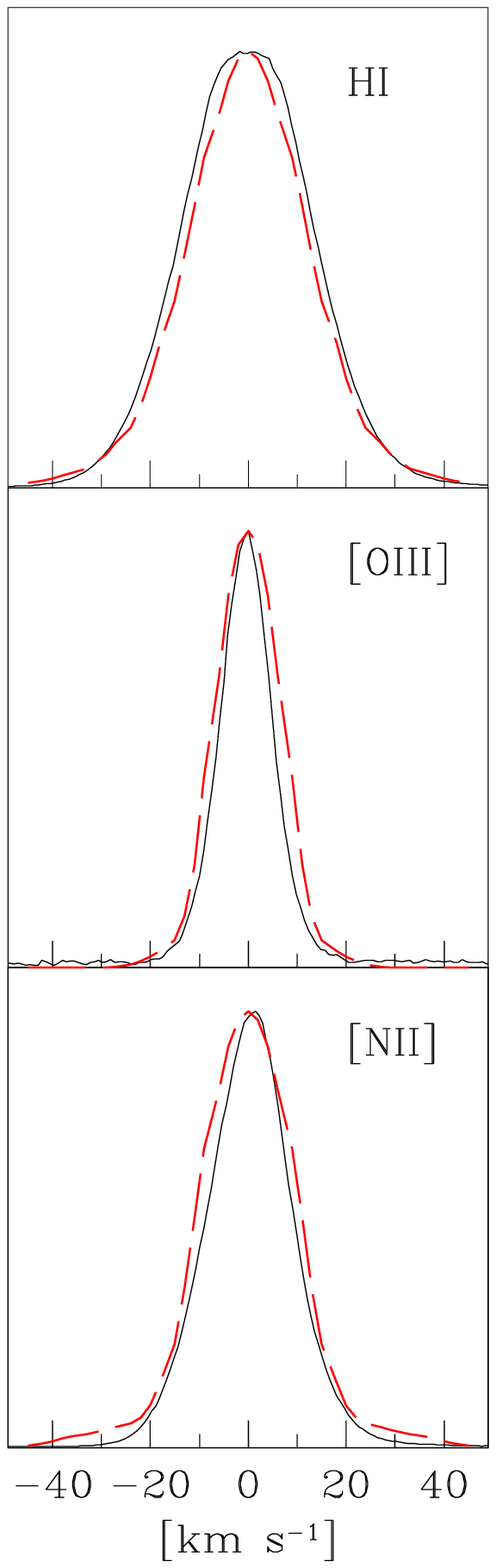} }
 \caption{Hen\,2-260.  Data presented as in
   Fig.\,\ref{mode0012}. The axis inclination is $50\degr$.
   The spectrograph slit (left middle panel) covers the whole nebula.}
 \label{mode0082}
\end{figure}

\begin{figure}
\resizebox{\hsize}{!}{
 \includegraphics[height=7cm]{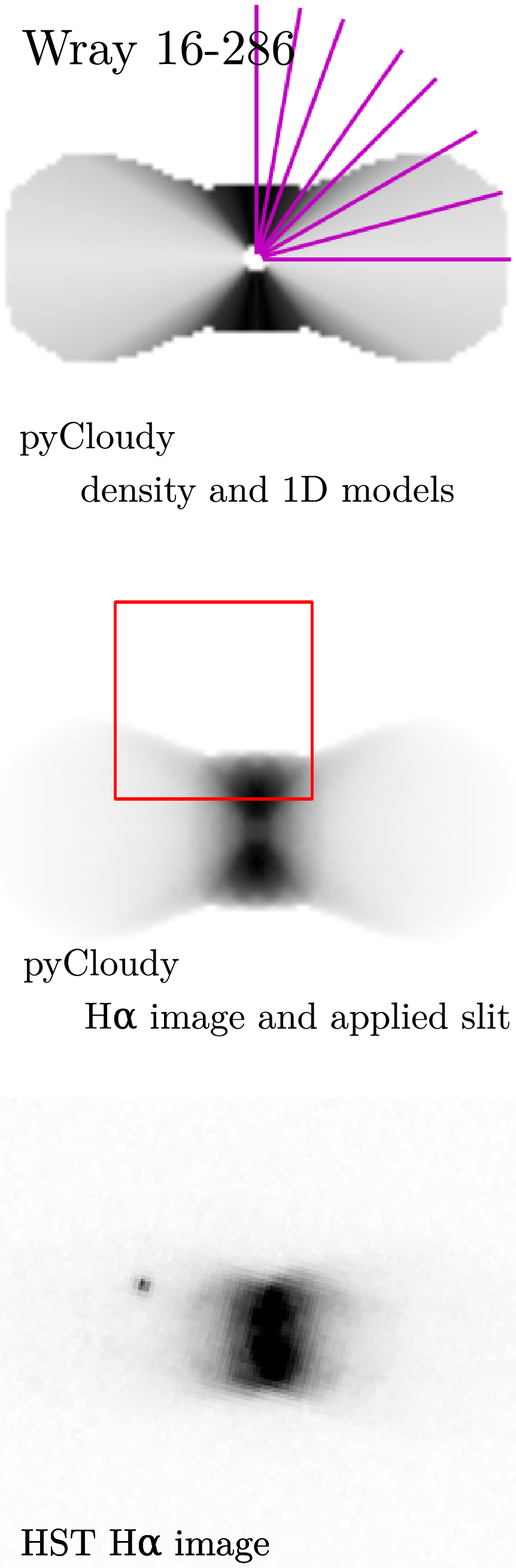}
 \includegraphics[height=7cm]{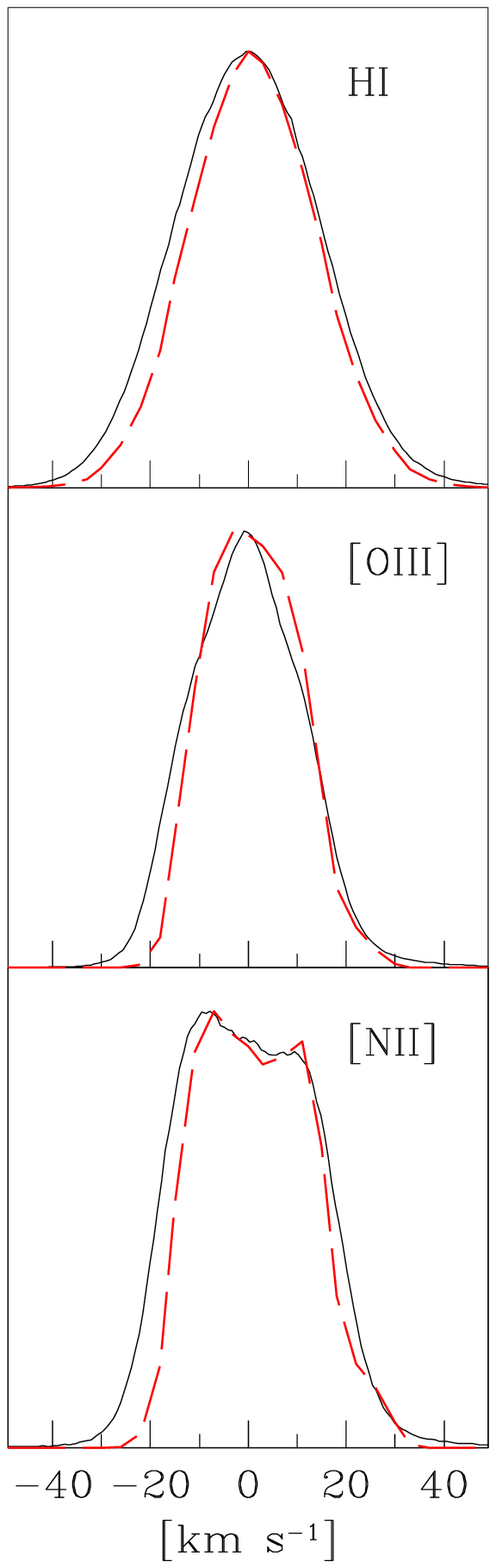} }
 \caption{Wray\,16-286. Data presented as in
   Fig.\,\ref{mode0012}. The axis inclination is $20\degr$.}
 \label{mode3519}
\end{figure}

\begin{figure}
\resizebox{\hsize}{!}{
 \includegraphics[height=7cm]{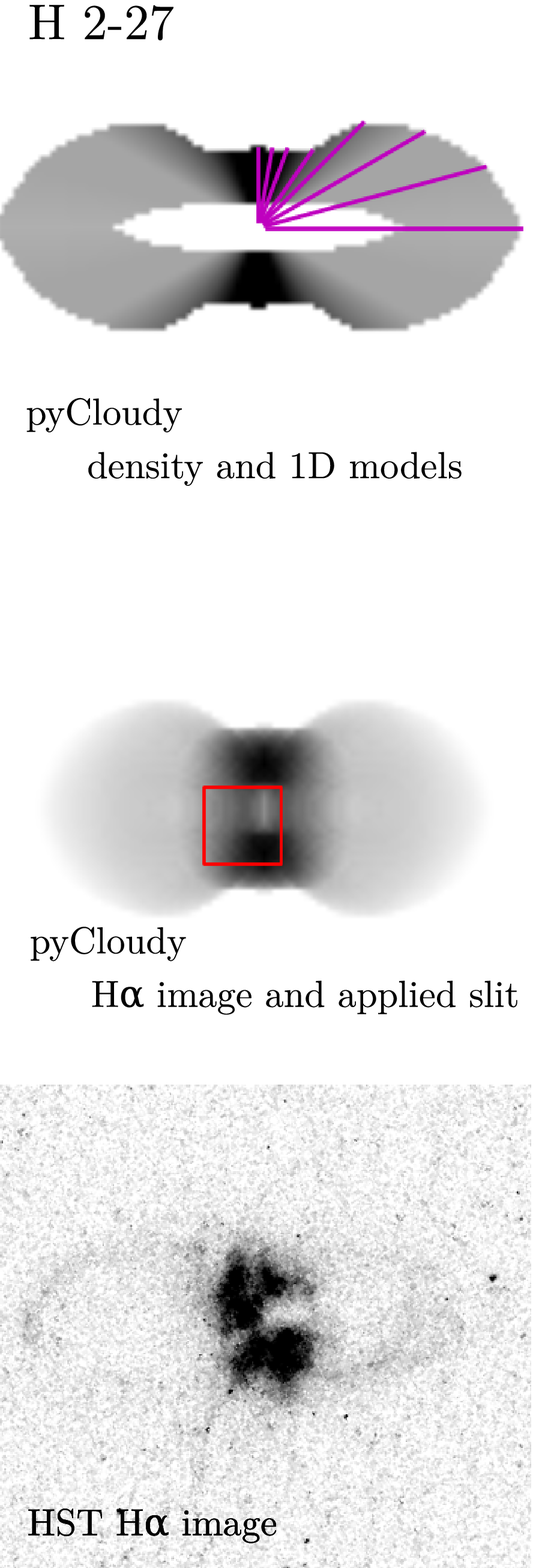}
 \includegraphics[height=7cm]{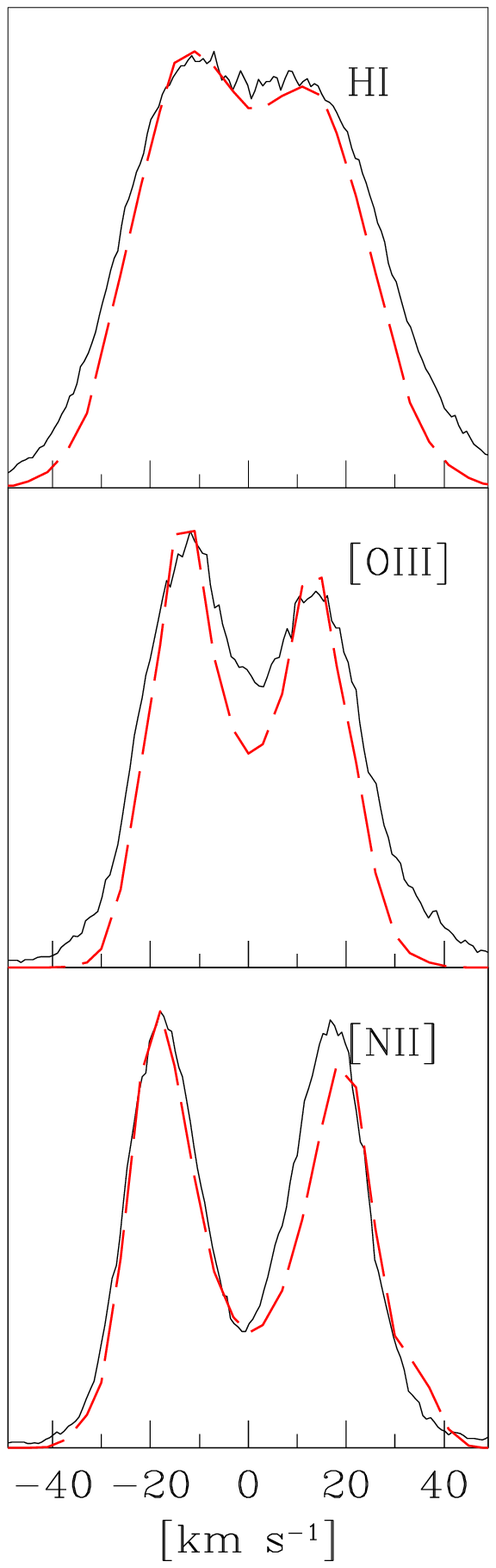} }
 \caption{H\,2-27. Data presented as in
   Fig.\,\ref{mode0012}. The axis inclination is $35\degr$. }
 \label{mode3565}
\end{figure}

\begin{figure}
\resizebox{\hsize}{!}{
 \includegraphics[height=7cm]{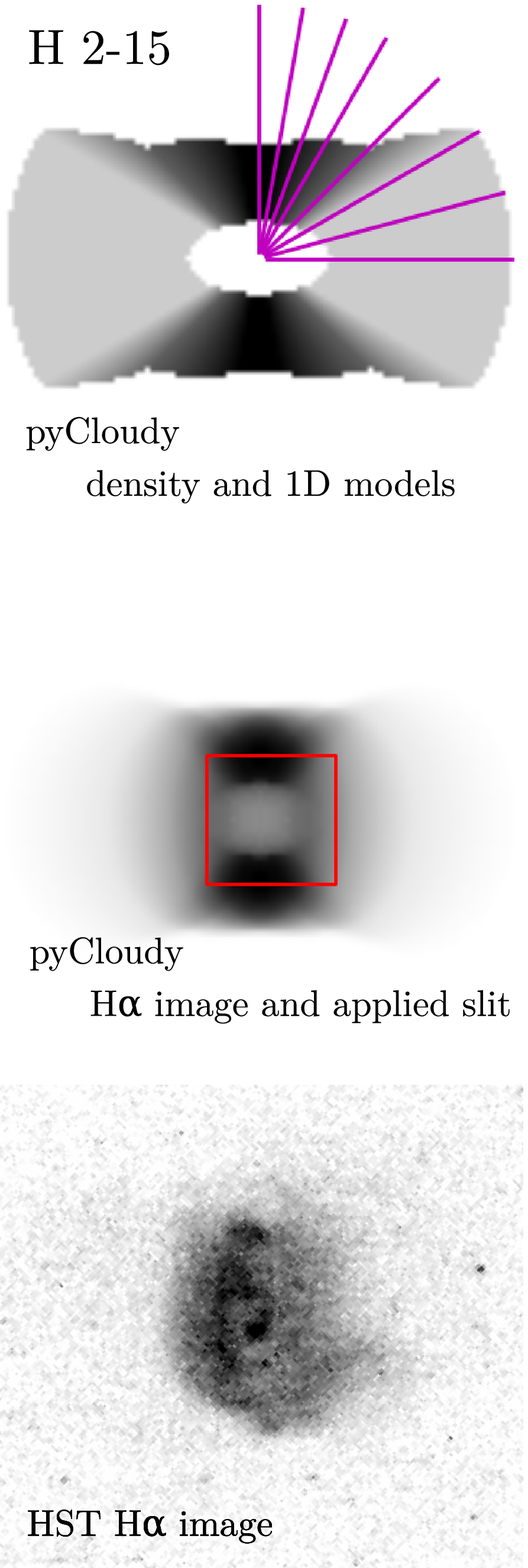}
 \includegraphics[height=7cm]{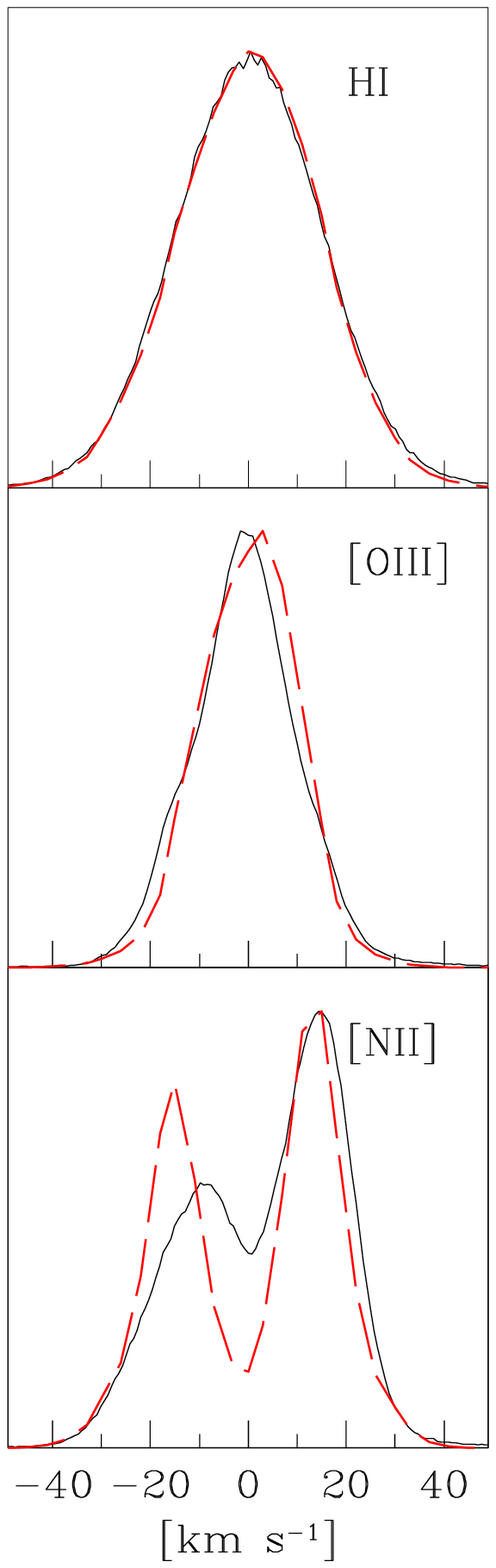} }
 \caption{H\,2-15. Data presented as in
   Fig.\,\ref{mode0012}. The axis inclination is $45\degr$. }
 \label{mode0038}
\end{figure}

\begin{figure}
\resizebox{\hsize}{!}{
 \includegraphics[height=7cm]{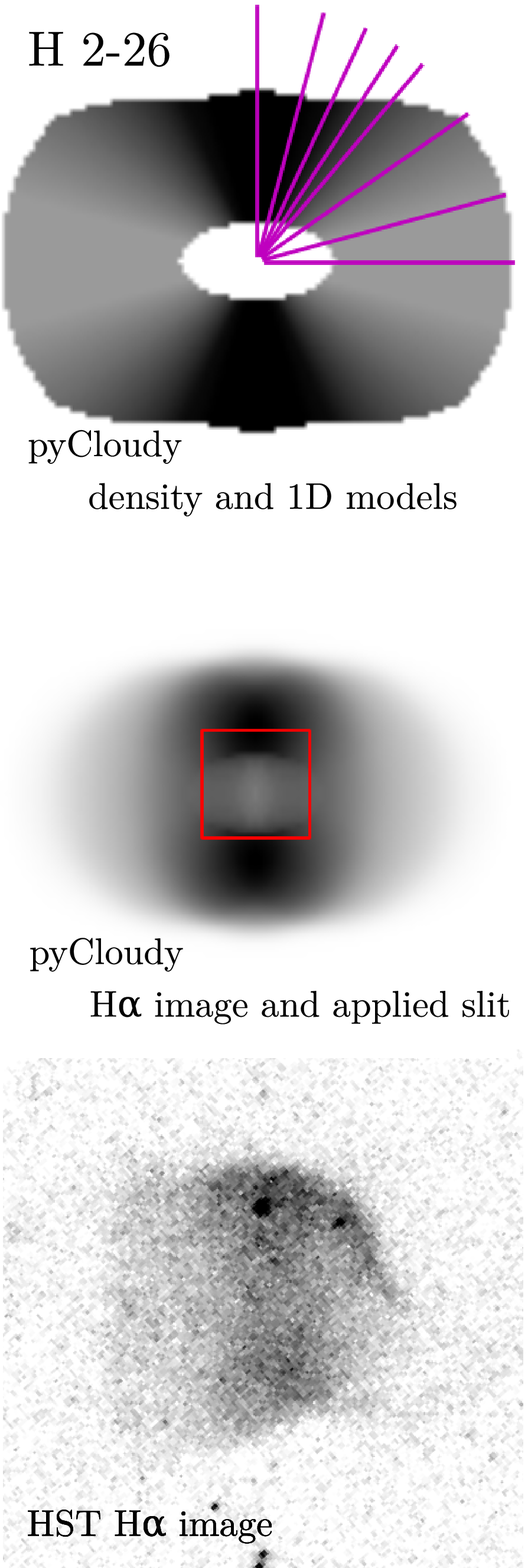}
 \includegraphics[height=7cm]{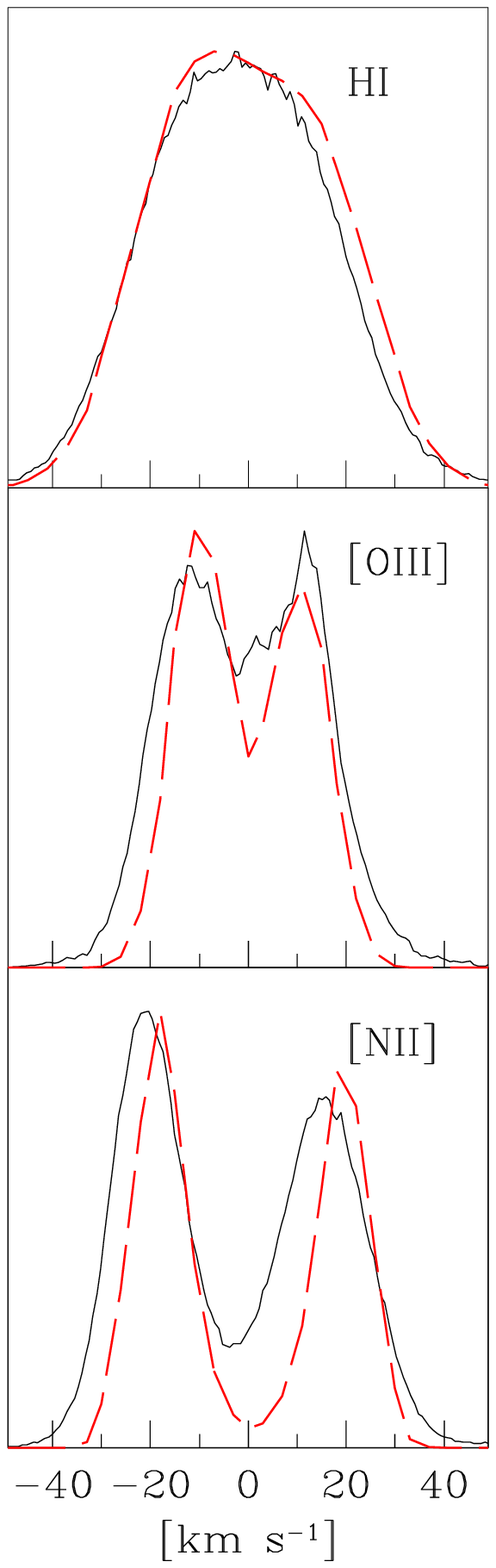} }
 \caption{H\,2-26. Data presented as in
   Fig.\,\ref{mode0012}. The axis inclination if $40\degr$.}
 \label{mode3561}
\end{figure}

Table\,\ref{param_photoio} lists the parameters of the
photo-ionization models for the program nebulae.  The Galactic
bulge distance was assumed by default however two objects discussed
below were considered lying outside the bulge. The stellar
blackbody temperature $T_\mathrm{bb}$ was adopted from previous Torun
models \citep{gesi2014}. We checked that major modifications were not
necessary to improve the fits, however we did allow for minor
corrections. The stellar luminosities were adopted to fit the nebular
H$\beta$ fluxes. In some cases these luminosities were smaller
than predicted by the evolutionary tracks what \citet{gesi2003}
interpreted in terms of photon leakage through transparent parts of the
nebulae.  In the axially symmetric objects the low-density and
low-opacity lobes are easily identified as responsible for such
leaking therefore the luminosity values should be regarded as lower
limits to the true stellar values.

For the line strengths and ratios we list the observed and the
pyCloudy values. The observed values are taken from the literature and
relate to the entire nebula.  The H$\beta$ fluxes are fitted well
because the densities and luminosities were varied to obtain that
fit. The two most representative strong lines are fitted with an
average error of about 25\% (in two extreme cases reaching 60\%)
which is actually not larger than in the old Torun models.  The
chemical abundances are taken from the Table\,7.2 (column
``Planetary'') of the Cloudy C13.1 documentation available at the web
page which assume solar-like metallicities. For N and O wherever
possible we changed the values to the available estimations for
individual PNe -- these data are given in the
Table\,\ref{param_photoio}.

The ionized mass is defined as the mass of the region where hydrogen
is ionized, and includes both the ionized and neutral helium in this
region.\footnote{pyCloudy returns the mass of ionized hydrogen. It
should be multiplied by a factor of 1.4 to obtain the mass of the
ionized region.}  For each object, two values of the ionized masses
are listed: the (1D) Torun and the pyCloudy model. The difference
between them is discussed in Section\,\ref{mass}.

The values of the inclination angles between the sky plane and
the nebula symmetry axis are presented in Table\,\ref{param_geom}
and in figure captions for each of the discussed objects. In
Figs.\,\ref{mode0012}--\ref{mode3561} we present the influence of the
inclination angle on the observed image: the upper left boxes show the
density distribution over the full length of the model while the
middle left boxes show the H$\alpha$ image obtained for the same model
but now with the long axis inclined to the plane of the sky. Depending
on the assumed angle the effect is more or less visible. These images
correspond to the (appropriately rotated) HST images shown in the
lower left boxes. This double presentation illustrates better the
intrinsic model structure versus the observed structure of the
nebula.

\subsection{\object{Hen 2-262} ( \object{PN G001.2+02.1})}
\label{hen2262}

The Torun model of Hen\,2-262 was not previously published and
therefore its parameters are not compared with pyCloudy in the
Table\,\ref{param_photoio}.  It was observed with the same set-up
together with the whole Galactic bulge sample however it was not
included in the analysis of \citet{gesi2014} because the complicated
velocity field required significant turbulent motions in the 1D
model. The pseudo-3D model shown in Fig.\,\ref{mode0012} is
acceptable, moving the slit off-centre improved the fit but it
still appears that adding some turbulence would broaden the line wings
and reduce the central splitting in better agreement with
observations. However, the presented model is with a simple, linear
velocity field.

The density contrast
between equator and pole is a factor of 5.6.  The model is radiation
bounded in the equatorial torus while the lobes are matter bounded.

\subsection{\object{H 2-20} ( \object{PN G002.8+01.7})}

The best fitting pyCloudy model (see Fig.\,\ref{mode0028}) has an
equator-to-pole density contrast of a factor of 5.  The model is
radiation bounded in all directions.

The emission line profiles show the presence of an ionization gradient
within the PN, which is usually associated with the presence of the
radiation boundary, i.e. ionization front. A small offset of the
artificial slit assumed at the line profile computations resulted in
asymmetric lines which resemble the ones observed. Nevertheless in
Fig.\,\ref{mode0028} more asymmetry can be seen in the observed
profiles which cannot be fitted with cylindrical-symmetry models.

The 1D model of \citet{gesi2014} predicted a velocity field which is
monotonically, almost linearly, increasing outwards with a small
plateau in the inner region. This is very similar to the velocity field
derived in this work.

\subsection{\object{H 2-18} ( \object{PN G006.3+04.4})}

The best-fitting model has a high equator-to-pole density contrast of a
factor of 11. The [\ion{N}{ii}]\,6583\AA\ line is very weak: any model with
an ionization front predicts this line an order of magnitude too
strong. The observed line ratios are best reproduced with a model which
is matter bounded and has low densities. However at these low
densities there is no ionization gradient within the nebula which is in
contradiction with the observed emission profiles. We built a
compromise model which is still matter bounded in all directions but
with a density of the equatorial torus high enough to show an
ionization gradient within the nebula.

The shape of the three emission lines is reproduced reasonably well,
including the unusual quadruple [\ion{O}{iii}]\,5007\AA\ line. However, the
details of [\ion{N}{ii}]\,6583\AA\ cannot be better tuned within the simple
model. The 1D model of \citet{gesi2014} predicted a linearly increasing
velocity field.

\subsection{\object{Hen 2-260} ( \object{PN G008.2+06.8})}

The equator-to-pole density contrast is a factor of 7. The best
fitting model (see Fig.\,\ref{mode0082}) shows a larger than average
inclination: this is required to reproduce the lack of splitting in
the [\ion{N}{ii}]\,6583\,\AA\ line profile. The nebula is radiation
bounded, except for a narrow cone along the symmetry axis where it is
matter bounded.

This object was discussed in \citet{hajd2014}, where it was analysed
with spherically symmetric (1D) Cloudy and Torun codes. They derived a
distance to this PN of 12\,kpc. We adopted this value, and also took
the other parameters from \citet{hajd2014}. The object was also in the
sample of \citet{gesi2014}, but there it was assumed to be at the
distance of Galactic bulge and therefore a smaller radius was used. The
low radial velocity of Hen\,2-260 \citep{durand1998} is consistent with
a non-bulge location.  Fig.\,\ref{hst_ima} shows that Hen\,2-260 is
much smaller than the other PNe which is understandable if it is
located beyond the Galactic bulge \citep{bensby2001}.

The line ratios are rather well reproduced by pyCloudy using a
blackbody $T_\mathrm{eff} = 26\,000$\,K. \citet{hajd2014} used a stellar
atmosphere model which requires a higher temperature for the same
number of ionizing photons: they find 35\,000\,K.  \citet{hajd2014}
derived the age of this PN with a direct method (from the heating rate
of the central star): their age (980\,yr) approximately agrees with that
estimated below from the kinematics.

The 1D model of \citet{gesi2014} predicted a parabolic shape for the
velocity field, with an average expansion velocity somewhat smaller
than the 3D estimation presented here. The linear velocity field used
here provides a reasonable fit to the line profiles, but there are
slight discrepancies in the widths of the various lines. Any further
improvement would require a more complex velocity field.

\subsection{\object{Wray 16-286} ( \object{PN G351.9-01.9})}

The equator-to-pole density contrast of the best model (see
Fig.\,\ref{mode3519}) is high, at a factor of 9.3.  The model is
radiation bounded in the equatorial torus whilst the lobes are matter
bounded.

The old 1D Torun models indicated a parabola-like velocity field with
mass-averaged velocity of 13\,km\,s$^{-1}$. This yields an age that
is significantly larger than pyCloudy: we found that the parabola
minimum is responsible for the difference.  The pyCloudy model
reproduces line profiles much better and the narrow
[\ion{O}{iii}]\,5007\,\AA\ and wide, flat-topped and slightly
asymmetric [\ion{N}{ii}]\,6583\,\AA\ are explained by the
spectrograph pointing off-centre with the slit not covering the
midpoint.  The pyCloudy model results in a higher average velocity of
around 20\,km\,s$^{-1}$ (dependent on the weighting method used in
the averaging).

\subsection{\object{H 2-27} ( \object{PN G356.5-03.6})}

The best fitting model (see Fig.\,\ref{mode3565}) gives a high
equator-to-pole density contrast, of a factor of 7.9.  The model
assumes a spectrograph slit position which is a little offset, in order
to reproduce the small asymmetry observed in profiles.  The model is
matter bounded in all directions.

For this PN we assumed a smaller distance than to the bulge. Although
reasonable photoionization models can be built for different distances,
some aspects point towards a smaller value. In Fig.\,\ref{hst_ima} this
PN is shown to have the largest angular dimensions. \citet{durand1998}
report a low radial velocity which does not require a bulge location. 
The correlations discussed below suggest that H\,2-27 has a smaller
kinematical age. The velocity is derived from spectra i.e.
independently from distance; therefore the age can be reduced by
decreasing the physical radius and therefore the distance. Having no
more constraints we simply assumed the distance of 4\,kpc for the
pyCloudy models. For H\,2-27 we recomputed the Torun model for the
smaller new distance, previous Torun models which assumed bulge distance
resulted in larger age for this PN.

The 1D model of \citet{gesi2014} predicted a linearly increasing
velocity field.

\subsection{Two post-bipolar nebulae: \object{H 2-15} ( \object{PN G003.8+05.3}) 
and \object{H 2-26} ( \object{PN G356.1-03.3}) }
\label{postbi1}

These two nebulae have a more irregular structure than the others, and
are suggested to be in a post-bipolar phase of evolution. They have 
numerous common characteristics and they are
included here for comparison. The models are shown in Figs.
\ref{mode0038} and \ref{mode3561}.

The models show a relatively low equator-to-density contrast of a
factor of 5 and 2.5 for H\,2-15 and H\,2-26 respectively. They have
lower density than the other objects peaking at around
$10^3$\,cm$^{-3}$.  Both models are composed of a thick disk which is
radiation bounded in the directions near the equatorial plane whilst
directions towards the poles are matter bounded.  For both, the
lowest stellar luminosities were derived so they can be the most
transparent for stellar ionizing radiation (most leaking).  Both
were fitted with linear velocity fields in \citet{gesi2014}. The 3D
pyCloudy resulted in a very similar averaged velocities. For these two
PNe \citet{gesi2014} built models with attached a low density halo to
the dense shell: this improved the fit of the
[\ion{N}{ii}]\,6583\AA\ line but in consequence increased the outer
radius and the age. The pyCloudy models assume for both nebulae that
the artificial spectrograph slit is a little offset, to reproduce the
asymmetry observed in profiles. However, the images also show
significant asymmetries.  The kinematical ages derived here are
smaller than those derived in \citet{gesi2014}, but are still two
times larger than the ages of the six bi-lobed PNe.

\section{Discussion}

\subsection{Ionized masses}
\label{mass}

For all objects the pyCloudy masses (presented in the
Table\,\ref{param_photoio}) are smaller than those from the 1D Torun
models, in most cases (6 out of 8) by a large factor of 2--7. In two
cases are the masses nearly the same.  The dominant effect is the use
of a 3D versus a 1D structure. A 3D model provides for a range of
density distributions within one object.

The H$\beta$ emission being proportional to the density squared
(actually $n_\mathrm{p} \times n_\mathrm{e}$) is dominated by
regions of high hydrogen density.  This gas has different spatial
distributions in the spherical and the pseudo-3D models, being a
compact sphere or a torus respectively.  Despite these differences the
densities and overall masses of those regions are comparable and
produce the same H$\beta$ fluxes.  Much larger are the differences in
the distributions of low density gas. In spherical models the low
density gas occupies the large volume of the spherical halo which sums
up to a significant fraction of the total nebular mass.

Another contributing factor is the irradiation of the outer regions. In
1D models, the radiation reaching the outer regions first passes
through dense inner layers and the attenuation gives a much reduced
intensity. To reach a significant line strength of the low-ionization
lines (which arise in the outer regions), a larger amount of gas is
required. In 3D models, there are directions in which the outer regions
are illuminated by radiation which has passed through little
intervening gas. Thus, less gas is needed to reproduce the line
strengths. 

This important result shows that ionized masses of planetary nebulae
have very large uncertainties attached. 1D models are best viewed as
giving upper limits to the mass. Pseudo-3D models provide a better 
approximation to true nebular ionized masses. 

The total mass lost by the central star might be somewhat larger than
the models ionized mass because we do not know how much neutral
material exists in the dense equatorial disk beyond the ionization
boundary.

\subsection{Averaged expansion velocities}

The true velocity distribution within a planetary nebula can in
principle be quite complicated.  The current analysis considers only
the large-scale flows which in our paper are only approximated, as
described in Sect.\,\ref{veloc}. Further simplification is related
to limiting the velocity to be a function of radius only, and not of
latitudinal angle.  Defining a single-valued expansion velocity for a
non-constant velocity field requires a weighting. The velocity at the
outer radius (or ionization front) is typically not useful, as it is
not directly measured: the available emission lines come from a region
a bit further in (e.g. the `post-shock' velocity discussed by
\citet{jaco2013}). For 1D models, the mass-averaged expansion velocity
\citep{gesi2000} has been shown to be a robust parameter. It
represents the bulk motion, and is easily and reproducibly measured
from the models, even where very few emission lines are available.

For the 3D models, there is not yet a similar methodology. We
investigate two options: weighting by mass (similar to the 1D models)
and weighting by the H$\beta$ emissivity. Weighting by mass has the
draw-back that the pyCloudy models use constant density (for each
direction). This gives a high weight to the outermost radius, where
velocities are highest but neither the densities nor the velocities
are well constrained.  The H$\beta$ emissivity decreases outwards 
tracing the ionized mass distribution.  Weighting by H$\beta$
emissivity gives smaller weight to the outer radius -- a behaviour
closer to the 1D Torun models, where the density decreased towards
the outer radius.

\begin{figure}
\centering
\includegraphics[width=6cm]{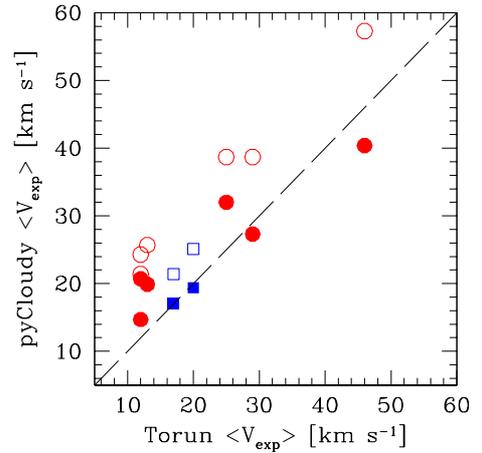}
      \caption{Averaged expansion velocity from Torun models compared
        with pyCloudy.  The abscissa shows the Torun models
        mass-averaged velocities, while at ordinate the pyCloudy
        H$\beta$-averaged values are shown as filled symbols, and the
        pyCloudy mass-averaged as open symbols. The (red) circles show
        the bipolar nebulae and the (blue) squares the post-bipolars.
        Data from Table \ref{param_photoio}.}
    \label{vON}
\end{figure}

Table\,\ref{param_photoio} lists the averaged expansion velocities
$<V_\mathrm{exp}>$ and nebular radii together with the corresponding
1D Torun model values.  Both averaging methods are compared in
  Fig.\,\ref{vON}, plotted versus old Torun model values.  Filled
  symbols represent weighting by H$\beta$ while open symbols weighting
  by mass distributions.  The circles (red) show the bipolar objects,
  the squares (blue) -- the probable post-bipolars.  The pyCloudy
mass-averaged velocities are higher than the H$\beta$-averaged
velocities -- exactly as expected. However note that in either case,
the result is dominated by the torus where most of the gas is located.

Comparing the values of $<V_\mathrm{exp}>$ from pyCloudy averaged by
H$\beta$ with the values of $<V_\mathrm{exp}>$ from Torun models averaged
by mass indicates a good correspondence (see Fig.\,\ref{vON}). 
We conclude that the averaged
velocity remains a robust parameter to describe a planetary nebula.
Because of the rather scarce observational and hydrodynamical
data to compare with, for the simplified models used here,
weighting by the H$\beta$ line emissivity is preferred over weighting
by mass.

\begin{figure}
\centering
\includegraphics[width=6cm]{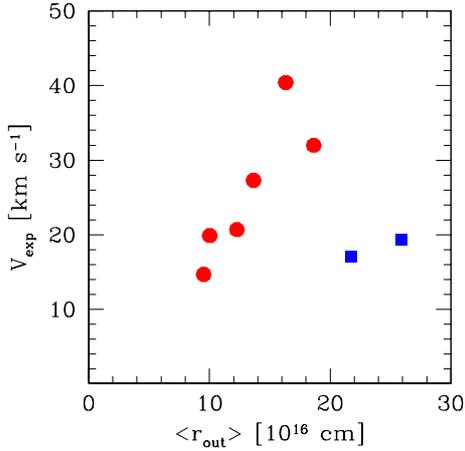}
      \caption{Averaged expansion velocity vs. averaged outer radius
        for the pyCloudy modelled PNe. Velocity are H$\beta$-averaged
        and the outer radius is mass averaged (see text). Data from
        Table \ref{param_photoio}. The filled circles show the bipolar nebulae
        and the filled squares the post-bipolars.}
    \label{fvr}
\end{figure}

In Fig.\,\ref{fvr} we plot the velocities versus the outer radius
radii. The velocities shown are H$\beta$ averaged wile the outer radii
are mass-weighted.  The circles represent the bipolar nebulae. They
show almost linear relation. The two PNe tentatively classified as
post-bipolar are represented by squares: they have the largest radii
but also show low expansion velocities.

\subsection{Kinematical ages}
\label{kinage}

The kinematical age can be derived from the (averaged) expansion velocity
and the outer radius. However, this only provides an approximation. The
ionization front travels faster than the gas velocity: the gas density
declines as the nebula expands, allowing more gas to become ionized.
The velocities can become very fast at the ionization front, due to the
over-pressure. To quantify the effects requires hydrodynamical models.
Based on the hydrodynamical models of \citet{Perinotto2004}, in the 1D
case the kinematical age should be reduced by a factor of 1.4
\citep{gesi2014}.

For 3D models the derivation is more complicated.  First, there is no unique
outer radius. Second, the torus and the lobes may also have different ages.
Third, the correction factors from 1D hydrodynamical models may not be
directly applicable to a non-spherical 3D structures, where pressure gradients
may cause non-radial velocities. Because of these uncertainties, we focus on
deriving limits to the ages.

An upper limit can be obtained from the outer radius and the expansion
velocity of the torus. We calculate the mass-averaged outer radius of
the ionized nebula. This is dominated by the dense torus; the value of
$<r_\mathrm{out}>$ is listed in the Table \ref{param_photoio}. We divide
this radius by the H$\beta$ weighted expansion velocity to obtain an
age. The H$\beta$-averaged value is used because it is smaller than
the mass-averaged value, and therefore yields a larger age. This age
is listed as `nebula' age, and should be considered an upper limit. No
correction factor other than the averaging has been applied. For
comparison, the 1D Torun model ages are also listed: for these the
correction factor of 1.4 has been used \citep{gesi2014, scho2005}.

A lower limit can be obtained from the lobes, which are expected to be younger
than the main nebula \citep{huggins2007}.  The linear velocity field indicates
a Hubble-type flow in the lobes. 
The inherent feature of homologous expansion is the same age
derived at any distance therefore the exact estimation of the extent of the
lobes is not an issue.
We assume that no acceleration or
deceleration has taken place and obtain the age for the tip of the
lobes. These ages are listed as `lobe' age in the Table \ref{param_photoio}.

Our `nebular' age is different from the Hubble-type age of the lobes.
The homologous expansion does not start from zero at the centre:
there is assumed a dense and constant-velocity inner region and the applied 
averaging is biased towards this low-velocity region.
Encouragingly, the upper limits are always higher than the lower
limits. The upper limits are in most cases close to the 1D Torun ages.

\subsection{Evolution}

\begin{figure}
\centering
\includegraphics[width=6cm]{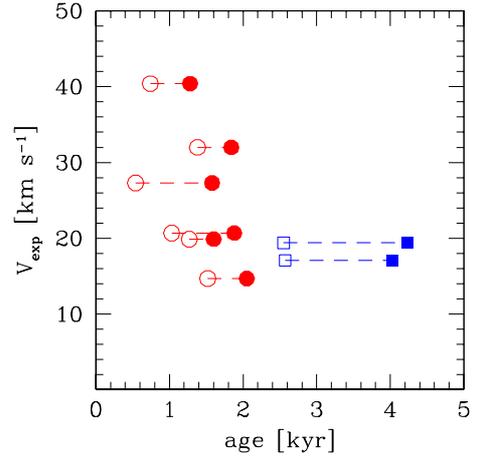}
      \caption{Averaged expansion velocity vs. kinematical age for the
        pyCloudy modelled PNe. The filled symbols show the nebular ages
        (derived from radii and velocities shown in Fig.\,\ref{fvr}),
        and the open symbols the lobe ages (see Sect.\,\ref{kinage}). Circles show the
        bipolars and squares the post-bipolars.}
    \label{fva}
\end{figure}

The aim of this work was to test for possible evolutionary sequences
among our objects. Fig.\,\ref{fva} shows the expansion velocity versus
the kinematical age. The open symbols show the ages of the lobes, and
the filled symbols the nebular ages. The latter is defined as above
and may be considered as upper limits.  All bipolar PNe occupy a
narrow region in age, in the range 1--2\,kyr. Some of the lobes are a
bit less than 1\,kyr. The nebular ages are 0.5--1\,kyr higher than the
lobes. For the two post-bipolars, the lobes are 2.5\,kyr old; their
nebular ages are around 4\,kyr, a much larger difference than for the
bipolars.

\subsubsection{The six bipolar PNe}

The interesting result is that the
selected bipolar PNe have nearly the same age (see Fig.\,\ref{fva}).
However their central star temperatures are spread over a wide range.
This may indicate a range of stellar masses (higher mass stars evolve
faster to higher temperature). This result suggests that the lobes may
form at a well-defined point during the nebular expansion, independent
of stellar mass.

None of the selected bipolars are older than 1.5--2\,kyr.  A sample of
six objects is too small to draw any general conclusions, but this may
point at a fading time for the lobes. A short window of visibility for
bipolar lobes would have consequences for their rate of occurrence.
The visibility time of the nebulae within the size range selected for
HST (5 arcsec or less) is 3--4\,kyr \citep[][their Fig.  11]{gesi2014}
(assuming the accelerated Sch\"onberner tracks). This is a
general visibility time which does not depend on the nebular
structure.  Our sample of 6 represents 17\%\ of the sample of
compact PNe observed with HST. The lobes can be seen for about half
of the total visibility period.  Correcting for this, we find that
30\%--35\%\ of the compact bulge PNe may experience a bipolar
phase. This number can be higher if it would be possible to
identify the bi-lobed PNe seen along the symmetry axis.

Our conclusion concerning ages is not unique among related objects.
\citet{huggins2007} has shown that jets and tori in post-AGB stars
have very similar ages, with jets being younger by a few hundred
years. The expansion times range from a couple of hundreds to a couple
of thousands years and the torus pre-dates the jet. This was also
found for the torus and lobes of NGC\,6302 \citep{szys2011}, where the
lobes are younger by $\sim$600\,yr from the 2900\,yr old torus. The
pre-PN M\,1-92 \citep{alco2007} kinematical age of 1200 yr (the same
in polar and equatorial directions of a bipolar molecular outflow) is
also very similar to our data.

\subsubsection{The two post-bipolar PNe}
\label{postbi}

In the sample of 36 PNe we tried to identify possible descendants of
the bipolar PNe. We searched amongst the oldest from the data of
\citet{gesi2014}. Two objects seem to be the likely candidates, namely
H\,2-15 and H\,2-26. Both are older than 3.5\,kyr (among the four
oldest), both have very hot central stars (supporting their
evolutionary advanced status), and both show morphologies resembling a
torus, already well expanded, with evidence for fragmentation. They also
reveal interactions with the interstellar medium. Some faint structures
might be the remnants of bipolar lobes. The derived velocity fields
contain a linear part, consistent with residual lobe emission. However,
more detailed observations would be needed to confirm this.

For both PNe the pyCloudy models confirmed their old kinematical age
although the new ages are somewhat smaller.  Both are larger than the
classic bipolar PNe (Fig.\,\ref{fvr}) and approximately two times
older (Fig.\,\ref{fva}).  

Their expansion velocities are low.
The velocities may be biased due to the weakness of the
lobes. However it is not unlikely that these two are descendants of
the subgroup of slow expanding PNe, while the objects which expand
faster after a couple of thousand years evolve beyond detection
limit. But this could be as well a result of the limited sample and
we just see such two objects. 

Some hint concerning the evolved nature and the timescales of
similarly looking PNe can be found in \citet{manc2015} where a
sequence of hydrodynamical models has been presented, intended to
explain the clumpy structure observed in PN NGC\,2346. The published
cross-sections through the nebular emission measures (unfortunately
not the rendered images) show how the equatorial toroid breaks up into
clumps already about 1000\,yr after the decline of the fast stellar
wind which earlier produced the bipolar lobes.  That is
interestingly similar to our post-bipolar objects.

It cannot be proven that the two PNe form late stages of bipolar
evolution. But in view of the short visibility time of bipolar lobes,
the presence of such objects in the bulge sample would be
expected. 
As noted in the previous subsection the phase of
symmetric lobes lasts about half of the total visibility time, so
the other half can appear like these.

\section{Conclusions}

\begin{itemize}

\item We selected targets from the HST survey of 36 compact bulge PNe, 
  not constrained on brightness or morphology.
  Six PNe in the sample have clear bi-lobed, axi-symmetric structure
  suitable for pseudo-3D modelling. Two further targets 
  are possibly in a post-bipolar phase. Most of the targets
  are believed to be at the distance of the Bulge, at 8\,kpc, 
  two of them are modelled at distances of 12\,kpc and 4\,kpc.

\item The pyCloudy models are able to fit the available constraints.
  The constant density per segment, and the simple velocity field,
  limit the accuracy but also keep the number of free parameters
  reasonable. 

\item The 3D models yield much lower ionized masses than the 1D
  models, typically reduced by a factor of 2--7. We find ionized
  masses in the range 0.02--0.1\,$M_\sun$. This is an important result
  with implications for the evolution of planetary nebulae.

\item The spectra are successfully reproduced with a linear
  radius-velocity relation in the outer regions. This simple modelling
  recipe appears in agreement with hydrodynamical calculations of
  \citet{huar2012} and with a very detailed analysis of the bipolar PN
  NGC\,6302 by \citet{szys2011}.  We caution that the current models
  did not explore a wide range of velocity fields, and the spectra
  were extracted at the central position of the nebula.

\item Averaged expansion velocities derived from the pyCloudy models are similar
  but not identical to the 1D models. Weighting the
  velocity field by the H$\beta$ line emissivity is preferred over
  weighting by mass, due to the limitations on the density distribution
  in this work. The average expansion velocity ranges from 15 to 30
  km\,s$^{-1}$, with one object (H\,2-18) expanding at 40\,km\,s$^{-1}$.

\item We derive ages for the lobes based on the model Hubble flow
  ranging from 500 to 1500\,yr. The two post-bipolar show older lobe
  ages, of 2500\,yr. The lobe ages provide lower limits to the age of
  the nebula. We derive upper limits to this age from the average outer
  radius and expansion velocity: these range from 1300 to 2000\,yr for
  the bipolars and 4000--4500 for the two post-bipolars. 

\item The age range of the bipolars is remarkably narrow. 
  The suggestion is that the bright lobes last a relatively
  short time, and fade after $\sim1000$\,yr. 

\item We cannot be confident that the two 'post-bipolar' PNe actual
  are descendants of bi-lobed objects. However, the derived parameters
  do not contradict such a relation.  A long known hypothesis
    recently summarized by \citet{frew2012} states that ``several
  stellar evolutionary pathways form objects classified as PNe''. Our
  morphological selection appears to have picked one such pathway,
  within which the clear bi-lobed structure can be seen at an early
  phase at ages of 1--2\,kyr while the post-bipolar phase can be
  identified at ages about 3\,kyr, all within the typical total
  visibility time of about 3--4\,kyr.  Still we do not know
    whether this pathway concerns single or binary star evolution.

\end{itemize}

\begin{acknowledgements}
This publication was supported by the Polish National Science Centre
(NCN) through the grant 2011/03/B/ST9/02552. We acknowledge the
Mexican project that made pyCloudy possible: CONACyT grant
CB-2010/153985.
\end{acknowledgements}

\begin{appendix}

\section{The photoionization model parameters and results}

\begin{table*}
\caption{Photoionization models parameters and modelling results for the program
  nebulae. The Galactic bulge distances were assumed by default however for two cases there were reasons to locate them outside the bulge. The temperature and luminosity correspond to the models central blackbody ionizing source. The extinction constants, line fluxes and abundances were taken from literature with the references indicated. Lower part of the table shows the resulting from computations ionized masses, averaged expansion velocities and kinematical ages. For a comparison (discussed in the text) there are given the new values obtained with pyCloudy code and the old values from the published Torun models. The expansion velocity parameter is weighted by H$\beta$ emission and by mass distribution, the old Torun values are only averaged by mass. The nebular outer radius is averaged by mass. The kinematical ages are: `lobe' -- the outer lobe radius divided by the velocity at this point, `nebula' -- the mass averaged outer radius divided by H$\beta$ averaged velocity. }
\begin{flushleft}
\begin{tabular}{lrrrrrrrr}
\hline\hline
\noalign{\smallskip}
PN\,G\,No. & 001.2+02.1 & 002.8+01.7 & 003.8+05.3 & 006.3+04.4 & 008.2+06.8 & 351.9-01.9 & 356.1-03.3 & 356.5-03.6 \\
PN name & \object{Hen 2-262} & \object{H 2-20} & \object{H 2-15} & \object{H 2-18} & \object{Hen 2-260} & \object{Wray 16-286} & \object{H 2-26} & \object{H 2-27} \\
\noalign{\smallskip}
\hline
\noalign{\smallskip}
Dist [kpc] & 8 & 8 & 8 & 8 & 12 & 8 & 8 & 4 \\
$T_\mathrm{bb}$ [K] & 69\,600 & 31\,600 & 80\,000 & 52\,000 & 26\,000 & 83\,000 & 152\,000 & 70\,000 \\
$L_*$  [$L_\sun$] & 2040 & 4070 & 130 & 1660 & 8130 & 7240 & 51  & 100 \\
\noalign{\smallskip}
\multicolumn{4}{l}{extinction constant} & \multicolumn{5}{}{} \\
 ~~$C_{\mathrm{H}\beta}$ & 2.7$^8$ & 2.5$^5$ & 1.0$^8$ & 1.63$^4$ & 0.69$^6$ & 2.5$^8$ & 1.0$^8$ & 2.2$^8$ \\
\noalign{\smallskip}

\multicolumn{4}{l}{$\log F(\mathrm{H}\beta)$ corrected} & \multicolumn{5}{}{} \\
~~observed & $-11.0^1$ & $-11.0^1$ & $-12.3^1$ & $-11.5^1$ & $-11.4^6$ & $-10.5^1$ & $-12.7^1$ & $-11.9^1$ \\
~~pyCloudy & $-11.0$ & $-11.1$ & $-12.2$ & $-11.6$ & $-11.3$ & $-10.5$ & $-12.7$ & $-12.0$ \\
\noalign{\smallskip}

\multicolumn{4}{l}{[\ion{N}{ii}]6583/H$\beta$ dered.} & \multicolumn{5}{}{} \\
~~observed &  $0.33^3$ & $2.81^7$ & $5.88^7$ & $0.08^4$ & $1.06^6$ & $0.59^1$ & $6.21^7$ & $1.58^1$ \\
~~pyCloudy &  $0.63$ & $3.50$ & $3.50$ & $0.07$ & $0.89$ & $0.61$ & $7.33$ & $1.60$ \\
\noalign{\smallskip}
\multicolumn{4}{l}{[\ion{O}{iii}]5007/H$\beta$ dered.} & \multicolumn{5}{}{} \\
~~observed &  $6.99^3$ & $0.27^7$ & $9.24^7$ & $13.00^4$ & $0.07^6$ & $10.60^1$ & $6.28^7$ & $7.01^1$ \\
~~pyCloudy &  $7.33$ & $0.42$ & $6.02$ & $5.24$ & $0.15$ & $14.30$ & $4.60$ & $5.58$ \\
\noalign{\smallskip}

\multicolumn{4}{l}{chemical abundances} & \multicolumn{5}{}{} \\
 ~~$\log ( \mathrm{N} / \mathrm{H} ) $ & $-4.18^3$ & $-3.40^7$ & -- & $-4.21^4$ & $-4.35^6$ & $-3.73^2$ & -- & -- \\
 ~~$\log ( \mathrm{O} / \mathrm{H} ) $ & $-3.52^3$ & $-2.74^7$ & $-3.34^7$ & $-3.44^4$ & $-3.31^6$ & $-3.26^2$ & $-3.66^7$ & -- \\

\noalign{\smallskip}
\hline
\noalign{\smallskip}

\multicolumn{4}{l}{ionized mass [$M_\odot$] } & \multicolumn{5}{}{} \\
~~Torun & -- & 0.11 & 0.15 &  0.27 &  0.05 &  0.25  & 0.14 &  0.02 \\
~~pyCloudy &  0.057 & 0.053 & 0.039 & 0.051 & 0.038 & 0.097 & 0.027 & 0.019 \\
\noalign{\smallskip}

\multicolumn{4}{l}{$<V_\mathrm{exp}>$ [km\,s$^{-1}$] averaged by} & \multicolumn{5}{}{} \\
~~H$\beta$ em. & 20.7  & 27.3 & 17.1 & 40.4  & 14.7  & 19.9  & 19.4  & 32.0  \\
~~mass         & 24.3 &  38.7 &  21.4 &  57.3 &  21.4 &  25.7 &  25.1 &  38.7 \\
\noalign{\smallskip}
\multicolumn{4}{l}{$<V_\mathrm{exp}>$ Torun [km\,s$^{-1}$]} & \multicolumn{5}{}{} \\
 &   --  &  29 &  17 &  46 &  13 &  13 &  20 &  25 \\
\noalign{\smallskip}
\multicolumn{4}{l}{averaged $<r_\mathrm{out}>$ [$10^{16}$\,cm] } & \multicolumn{5}{}{} \\
 &  12.28 & 13.65 & 21.77 & 16.32 & 9.52 & 10.03 & 25.92 & 18.64 \\
\noalign{\smallskip}
\multicolumn{4}{l}{kinematical age [kyr]} & \multicolumn{5}{}{} \\
~~Torun     & --   & 1.56 & 4.93 & 1.67 & 1.61 & 3.22 & 3.84 & 1.63 \\
~~nebula    & 1.88 & 1.58 & 4.03 & 1.28 & 2.05 & 1.60 & 4.23 & 1.84 \\
~~lobe      & 1.03 & 0.54 & 2.57 & 0.74 & 1.52 & 1.27 & 2.55 & 1.38 \\
\noalign{\smallskip}
\hline
\noalign{\smallskip}
\multicolumn{9}{l}{References: ~~
$^1$ \citet{acker1992}; $^2$ \citet{chia2009}; $^3$ \citet{escu2004}; $^4$ \citet{GCSC2009}; }\\
\multicolumn{9}{l}{ ~~~~~~~~~~~~~~~
$^5$ \citet{gute2008}; $^6$ \citet{hajd2014}; $^7$ \citet{stas1998}; $^8$ \citet{TASK1992}. }\\
\noalign{\smallskip}
\hline
\end{tabular}
\end{flushleft}
\label{param_photoio}
\end{table*}


\begin{table*}
\caption{Geometry defining parameters for the program nebulae. The first line shows the inclination angle between the PN long axis and the plane of the sky. Then follow eight groups of lines presenting parameters of each of latitudinal segments out of which the pseudo-3D models are interpolated. The inner and outer radii are in the units of [10$^{16}$\,cm], the hydrogen number density is in units of [cm$^{-3}$].  }
\begin{flushleft}
\begin{tabular}{lrrrrrrrr}
\hline\hline
\noalign{\smallskip}
PN\,G\,No. & 001.2+02.1 & 002.8+01.7 & 003.8+05.3 & 006.3+04.4 & 008.2+06.8 & 351.9-01.9 & 356.1-03.3 & 356.5-03.6 \\
PN name & \object{Hen 2-262} & \object{H 2-20} & \object{H 2-15} & \object{H 2-18} & \object{Hen 2-260} & \object{Wray 16-286} & \object{H 2-26} & \object{H 2-27} \\
\noalign{\smallskip}
\hline
\noalign{\smallskip}
inclination & 20$\degr$ & 30$\degr$ & 45$\degr$ & 35$\degr$ & 50$\degr$ & 20$\degr$ & 40$\degr$ & 35$\degr$ \\
\noalign{\smallskip}
latitude            & 0$\degr$$\qquad$ & 0$\degr$$\qquad$ & 0$\degr$$\qquad$ & 0$\degr$$\qquad$ & 0$\degr$$\qquad$ & 0$\degr$$\qquad$ & 0$\degr$$\qquad$ & 0$\degr$$\qquad$ \\
$\quad$$r_\mathrm{in}$  & 8.9 & 8.5 & 5.0 & 2.0 & 2.6 & 1.0 & 5.0 & 3.2 \\
$\quad$$r_\mathrm{out}$  & 10.8 & 10.5 & 16.1 & 11.7 & 5.5 & 6.0 & 21.9 & 10.5 \\
$\quad$$\log n_\mathrm{H}$  & 4.2 & 4.1 & 3.2 & 3.65 & 4.5 & 4.7 & 2.9 & 3.15 \\
\noalign{\smallskip}
latitude             & 15$\degr$$\qquad$  &  15$\degr$$\qquad$ & 10$\degr$$\qquad$  & 15$\degr$$\qquad$  & 15$\degr$$\qquad$  & 10$\degr$$\qquad$  & 15$\degr$$\qquad$  & 10$\degr$$\qquad$  \\
$\quad$$r_\mathrm{in}$  & 8.4  & 8.5  & 5.1  & 2.1  & 2.8  &  1.0 & 5.1  &  3.2 \\
$\quad$$r_\mathrm{out}$& 10.5  & 10.4  & 16.1  & 12.0  & 5.7  & 5.7  & 21.9  & 10.5  \\
$\quad$$\log n_\mathrm{H}$  & 4.2  & 4.1  &  3.2 &  3.65  & 4.4  & 4.7  & 2.9 & 3.15 \\
\noalign{\smallskip}
latitude             & 30$\degr$ $\qquad$ & 30$\degr$ $\qquad$ & 20$\degr$ $\qquad$ & 30$\degr$$\qquad$  & 30$\degr$ $\qquad$ & 20$\degr$ $\qquad$ & 25$\degr$ $\qquad$ &  20$\degr$ $\qquad$\\
$\quad$$r_\mathrm{in}$ & 7.4  & 8.3  & 5.2  & 2.3  & 3.1  &  1.0 & 5.4  & 3.4  \\
$\quad$$r_\mathrm{out}$ & 9.9  & 11.2  & 17.4  & 12.3  & 6.1  & 6.2  & 22.7  &  10.9 \\
$\quad$$\log n_\mathrm{H}$  & 4.2  &  4.0 & 3.2  &  3.65  & 4.4  & 4.8  & 2.9  & 3.15 \\
\noalign{\smallskip}
latitude             & 45$\degr$ $\qquad$ & 40$\degr$ $\qquad$ & 30$\degr$ $\qquad$ & 45$\degr$ $\qquad$ & 45$\degr$$\qquad$  & 35$\degr$$\qquad$  & 3$\degr$3 $\qquad$ & 35$\degr$$\qquad$  \\
$\quad$$r_\mathrm{in}$  & 6.5  & 8.2  &  5.6 & 2.7  & 3.7  & 1.0  & 5.7  & 3.8  \\
$\quad$$r_\mathrm{out}$ &10.5   & 11.6  & 18.8  & 12.6  & 7.0  & 6.7  & 25.0  &  12.3 \\
$\quad$$\log n_\mathrm{H}$ & 4.1  & 4.0  & 3.1  & 3.45  & 4.3  & 4.6  & 2.8  & 3.0  \\
\noalign{\smallskip}
latitude             & 55$\degr$ $\qquad$ & 50$\degr$ $\qquad$ & 45$\degr$ $\qquad$ & 55$\degr$$\qquad$  & 60$\degr$$\qquad$  & 45$\degr$$\qquad$  & 40$\degr$ $\qquad$ & 45$\degr$ $\qquad$ \\
$\quad$$r_\mathrm{in}$  & 6.0  & 8.1  & 6.3  & 3.3  & 4.9  & 1.0  & 6.0  &  4.4 \\
$\quad$$r_\mathrm{out}$ & 13.4  & 13.7  & 22.0  & 20.0  & 9.4  & 9.0  & 27.0  & 19.5  \\
$\quad$$\log n_\mathrm{H}$  & 3.9  & 3.8  & 3.0  & 3.2  & 4.1  & 4.4  & 2.8  & 2.9  \\
\noalign{\smallskip}
latitude           & 65$\degr$ $\qquad$ & 60$\degr$$\qquad$  & 60$\degr$ $\qquad$ & 65$\degr$ $\qquad$ & 75$\degr$ $\qquad$ &  60$\degr$ $\qquad$& 55$\degr$ $\qquad$ &  60$\degr$$\qquad$ \\
$\quad$$r_\mathrm{in}$ & 5.7  & 7.9  &  7.6 & 4.2  & 7.6  &  1.0 & 7.1  & 6.1  \\
$\quad$$r_\mathrm{out}$  & 16.4  & 17.7  & 35.4  & 31.6  & 25.1  & 16.7  & 32.1  & 25.1  \\
$\quad$$\log n_\mathrm{H}$  & 3.7  & 3.6  &  2.5 &3.0   & 3.4  &  4.0 & 2.7  &  2.7 \\
\noalign{\smallskip}
latitude            & 75$\degr$ $\qquad$ &  75$\degr$$\qquad$ & 75$\degr$$\qquad$  & 75$\degr$$\qquad$  & 85$\degr$$\qquad$  & 75$\degr$ $\qquad$ & 75$\degr$ $\qquad$ & 75$\degr$$\qquad$  \\
$\quad$$r_\mathrm{in}$  & 5.5  &  7.8 & 9.1  & 5.6  & 10.2  &  1.0 &9.1   &  10.4 \\
$\quad$$r_\mathrm{out}$  & 19.8  & 21.8  & 35.4  & 35.3  & 25.1  & 20.0  & 33.1  & 30.9  \\
$\quad$$\log n_\mathrm{H}$  & 3.6  & 3.5  & 2.5  & 2.6  & 3.4  & 3.9  & 2.5  &  2.7 \\
\noalign{\smallskip}
latitude             & 90$\degr$  $\qquad$& 90$\degr$ $\qquad$ & 90$\degr$ $\qquad$ & 90$\degr$$\qquad$  & 90$\degr$$\qquad$  & 90$\degr$ $\qquad$ & 90$\degr$$\qquad$  & 90$\degr$$\qquad$  \\
$\quad$$r_\mathrm{in}$  &  5.3 &  7.8 & 10.0  & 8.0  & 10.8  & 1.0  & 10.0  &  19.0 \\
$\quad$$r_\mathrm{out}$  & 24.0  & 23.5  & 35.4  & 39.7  & 25.1  & 20.0  & 33.1  & 34.7  \\
$\quad$$\log n_\mathrm{H}$ & 3.5  & 3.4  &2.5   & 2.6  & 3.4  & 3.7  &2.5   & 2.65  \\
\noalign{\smallskip}
\hline
\end{tabular}
\end{flushleft}
\label{param_geom}
\end{table*}

\end{appendix}

\end{document}